\documentclass[12pt,epsf]{article}

\usepackage{times}
\usepackage{graphicx}
\usepackage{amsfonts}
\usepackage{amsmath}
\usepackage{amssymb}
\usepackage{amstext}
\usepackage{amsthm}

\usepackage{epsfig}
\psfigdriver{dvips}
\usepackage{latexsym}
\usepackage[matrix,curve]{xypic}
\usepackage{rotating}
\rotdriver{dvips}

\begin{document}

\baselineskip 6mm
\renewcommand{\thefootnote}{\fnsymbol{footnote}}


\newcommand{\nc}{\newcommand}
\newcommand{\rnc}{\renewcommand}


\rnc{\baselinestretch}{1.24}    
\setlength{\jot}{6pt}       
\rnc{\arraystretch}{1.24}   

\makeatletter
\rnc{\theequation}{\thesection.\arabic{equation}}
\@addtoreset{equation}{section}
\makeatother



\nc{\be}{\begin{equation}}

\nc{\ee}{\end{equation}}

\nc{\bea}{\begin{eqnarray}}

\nc{\eea}{\end{eqnarray}}

\nc{\xx}{\nonumber\\}

\nc{\ct}{\cite}

\nc{\la}{\label}

\nc{\eq}[1]{(\ref{#1})}

\nc{\newcaption}[1]{\centerline{\parbox{6in}{\caption{#1}}}}

\nc{\fig}[3]{

\begin{figure}
\centerline{\epsfxsize=#1\epsfbox{#2.eps}}
\newcaption{#3. \label{#2}}
\end{figure}
}


\def\CA{{\cal A}}
\def\CC{{\cal C}}
\def\CD{{\cal D}}
\def\CE{{\cal E}}
\def\CF{{\cal F}}
\def\CG{{\cal G}}
\def\CH{{\cal H}}
\def\CK{{\cal K}}
\def\CL{{\cal L}}
\def\CM{{\cal M}}
\def\CN{{\cal N}}
\def\CO{{\cal O}}
\def\CP{{\cal P}}
\def\CR{{\cal R}}
\def\CS{{\cal S}}
\def\CU{{\cal U}}
\def\CV{{\cal V}}
\def\CW{{\cal W}}
\def\CY{{\cal Y}}
\def\CZ{{\cal Z}}


\def\IB{{\hbox{{\rm I}\kern-.2em\hbox{\rm B}}}}
\def\IC{\,\,{\hbox{{\rm I}\kern-.50em\hbox{\bf C}}}}
\def\ID{{\hbox{{\rm I}\kern-.2em\hbox{\rm D}}}}
\def\IF{{\hbox{{\rm I}\kern-.2em\hbox{\rm F}}}}
\def\IH{{\hbox{{\rm I}\kern-.2em\hbox{\rm H}}}}
\def\IN{{\hbox{{\rm I}\kern-.2em\hbox{\rm N}}}}
\def\IP{{\hbox{{\rm I}\kern-.2em\hbox{\rm P}}}}
\def\IR{{\hbox{{\rm I}\kern-.2em\hbox{\rm R}}}}
\def\IZ{{\hbox{{\rm Z}\kern-.4em\hbox{\rm Z}}}}


\def\a{\alpha}
\def\b{\beta}
\def\d{\delta}
\def\ep{\epsilon}
\def\ga{\gamma}
\def\k{\kappa}
\def\l{\lambda}
\def\s{\sigma}
\def\t{\theta}
\def\w{\omega}
\def\G{\Gamma}


\def\half{\frac{1}{2}}
\def\dint#1#2{\int\limits_{#1}^{#2}}
\def\goto{\rightarrow}
\def\para{\parallel}
\def\brac#1{\langle #1 \rangle}
\def\curl{\nabla\times}
\def\div{\nabla\cdot}
\def\p{\partial}


\def\Tr{{\rm Tr}\,}
\def\det{{\rm det}}


\def\vare{\varepsilon}
\def\zbar{\bar{z}}
\def\wbar{\bar{w}}
\def\what#1{\widehat{#1}}


\def\ad{\dot{a}}
\def\bd{\dot{b}}
\def\cd{\dot{c}}
\def\dd{\dot{d}}
\def\so{SO(4)}
\def\bfr{{\bf R}}
\def\bfc{{\bf C}}
\def\bfz{{\bf Z}}

\begin{titlepage}


\hfill\parbox{3.7cm} {HU-EP-06/21 \\
{\tt hep-th/0611174}}

\vspace{15mm}

\begin{center}
{\Large \bf  Emergent Gravity from Noncommutative Spacetime}

\vspace{10mm}
Hyun Seok Yang \footnote{hsyang@physik.hu-berlin.de}
\\[10mm]

{\sl Institut f\"ur Physik, Humboldt Universit\"at zu Berlin \\
Newtonstra\ss e 15, D-12489 Berlin, Germany}

\end{center}

\thispagestyle{empty}

\vskip1cm


\centerline{\bf ABSTRACT}
\vskip 4mm
\noindent
We showed before that self-dual electromagnetism in noncommutative
(NC) spacetime is equivalent to self-dual Einstein gravity. This
result implies a striking picture about gravity: Gravity can emerge
from electromagnetism in NC spacetime. Gravity is then a collective
phenomenon emerging from gauge fields living in fuzzy spacetime. We
elucidate in some detail why electromagnetism in NC spacetime should
be a theory of gravity. In particular, we show that NC
electromagnetism is realized through the Darboux theorem as
a diffeomorphism symmetry $G$ which is spontaneously broken to
symplectomorphism $H$ due to a background
symplectic two-form $B_{\mu\nu}=(1/\theta)_{\mu\nu}$,
giving rise to NC spacetime. This leads to a natural speculation
that the emergent gravity from NC electromagnetism corresponds to a
nonlinear realization $G/H$ of the diffeomorphism group, more
generally its NC deformation. We also find some evidences that the
emergent gravity contains the structures of generalized complex
geometry and NC gravity. To illuminate the emergent gravity, we
illustrate how self-dual NC electromagnetism nicely fits with the
twistor space describing curved self-dual spacetime. We also discuss
derivative corrections of Seiberg-Witten map which give rise to
higher order gravity.
\\


\vspace{1cm}

\today

\end{titlepage}

\renewcommand{\thefootnote}{\arabic{footnote}}
\setcounter{footnote}{0}

\section{Introduction and Symplectic Geometry}

Recently we showed in \ct{mine} that self-dual electromagnetism in
noncommutative (NC) spacetime is equivalent to self-dual Einstein gravity.
For example, $U(1)$ instantons in NC spacetime are actually gravitational
instantons \ct{sty,ys}. This result implies a striking picture about gravity:
Gravity can emerge from electromagnetism if the spacetime, at
microscopic level, is noncommutative like the quantum mechanical
world. Gravity is then a collective phenomenon emerging from gauge
fields living in fuzzy spacetime. Similar ideas have been described
in \ct{nc-induced-gravity} and in a recent review \ct{szabo}
that NC gauge theory can naturally induce a gauge theory of gravitation.

In this paper we will show that the ``emergent gravity'' from NC spacetime is
very generic in NC field theories, not only restricted to the self-dual
sectors. Since this picture about gravity is rather unfamiliar,
though marked evidences from recent developments in string and M theories
are ubiquitous, it would be desirable to have an intuitive picture
for the emergent gravity. This remarkable physics turns out to be deeply
related to symplectic geometry in sharp contrast to Riemannian geometry.
Thus we first provide conceptual insights, based on intrinsic properties of
the symplectic geometry, on why a field theory formulated on NC spacetime
could be a theory of gravity. We will discuss more concrete
realizations in the coming sections. We refer \ct{books} for rigorous
details about the symplectic geometry.

{\bf Symplectic manifold}: A symplectic structure on a smooth manifold $M$ is
a non-degenerate, closed 2-form $\omega \in \Lambda^2(M)$. The pair
$(M,\omega)$ is called a symplectic manifold. In classical mechanics,
a basic symplectic manifold is the phase space of $N$-particle
system with $\omega = \sum  dq^i \wedge dp_i$.

A NC spacetime is obtained by introducing a symplectic structure $B = \half
B_{\mu\nu} dy^\mu \wedge dy^\nu$ and then by quantizing the spacetime with
its Poisson structure $\theta^{\mu\nu} \equiv (B^{-1})^{\mu\nu}$,
treating it as a quantum phase space. That is, for $f,g \in C^\infty(M)$,
\be \la{poisson-bracket}
\{f, g\} = \theta^{\mu\nu} \left(\frac{\p f}{\p y^\mu} \frac{\p g}{\p y^\nu}
- \frac{\p f}{\p y^\nu} \frac{\p g}{\p y^\mu}\right) \Rightarrow
-\frac{i}{\hbar}[\what{f},\what{g}],
\ee
where $\hbar$ is a formal parameter and we sometimes set $\hbar=1$ by
absorbing it in $\theta$.

According to the Weyl-Moyal map \ct{nc-review}, the NC algebra of operators
is equivalent to the deformed algebra of functions defined
by the Moyal $\star$-product, i.e.,
\begin{equation}\label{star-product}
\what{f} \cdot \what{g} \cong (f \star g)(y) = \left.\exp\left(\frac{i\hbar}{2}
\theta^{\mu\nu} \partial_{\mu}^{y}\partial_{\nu}^{z}\right)f(y)g(z)\right|_{y=z}.
\end{equation}

{\bf Symplectomorphism}: Let $(M,\omega)$ be a symplectic manifold.
Then a diffeomorphism $\phi: M \to M$ satisfying $\omega = \phi^*
(\omega)$ is a symplectomorphism. In classical mechanics, symplectomorphisms
are called canonical transformations. A vector field $X$ on $M$
is said to be symplectic if $\CL_X \omega = 0$. The Lie derivative along a
vector field $X$ satisfies the Cartan's homotopy formula $\CL_X = \iota_X d + d
\iota_X$, where $\iota_X$ is the inner product with $X$, e.g., $\iota_X \omega
(Y) = \omega(X,Y)$. Since $d\omega =0$, $X$ is a symplectic vector field
if and only if $\iota_X \omega$ is closed. If $\iota_X \omega$ is exact, i.e.,
$\iota_X \omega = d H$ for any $H \in C^\infty(M)$, $X$ is called the
Hamiltonian vector field. So the first cohomology group $H^1(M)$ measures the
obstruction for a symplectic vector field to be Hamiltonian. Since we are
interested in simply connected manifolds, e.g., $M=\IR^4$, every symplectic
vector field would be Hamiltonian.

Through the quantization rule \eq{poisson-bracket} and \eq{star-product},
one can define NC $\IR^4$ by the following commutation relation
\be \la{nc-spacetime}
[y^\mu, y^\nu]_\star = i \theta^{\mu\nu}.
\ee
An important point is that the set of diffeomorphisms generated
by Hamiltonian vector fields, denoted as $Ham(M)$, preserves the NC
algebra \eq{nc-spacetime} since $\CL_X B = 0$ with $B = \theta^{-1}$
provided $\iota_X B = d \lambda$ where $\lambda$ is an arbitrary
function \ct{cornalba,jackiw-pi-poly}. The symmetry $Ham(M)$ is
infinite-dimensional as well as non-Abelian and can be identified
with NC $U(1)$ gauge group \ct{cornalba} upon quantization in the sense
of Eq.\eq{poisson-bracket}. The NC gauge symmetry then acts as unitary
transformations on an infinite-dimensional, separable Hilbert space
$\mathcal{H}$ which is the representation space of the Heisenberg
algebra \eq{nc-spacetime}. This NC gauge symmetry $U_{\rm{cpt}}(\mathcal{H})$
is so large that $U_{\rm{cpt}}({\mathcal{H}}) \supset U(N)
\;(N \rightarrow \infty)$ \cite{harvey}. In this sense the NC gauge theory
is essentially a large $N$ gauge theory. It becomes more explicit on
a NC torus through the Morita equivalence where NC
$U(1)$ gauge theory with rational $\theta = M/N$ is equivalent to an ordinary
$U(N)$ gauge theory \ct{sw,morita}. Therefore it is not so surprising that NC
electromagnetism shares essential properties appearing in a large
$N$ gauge theory such as $SU(N \to \infty)$ Yang-Mills theory or matrix models.

The symplectic manifolds accompany an important property, the
so-called Darboux theorem, stating that every symplectic manifold is
locally symplectomorphic.

{\bf Darboux theorem}: Locally, $(M,\omega) \cong (\IC^n, \sum dq^i \wedge
dp_i).$  That is, every $2n$-dimensional symplectic manifold can always
be made to look locally like the linear symplectic space $\IC^n$ with
its canonical symplectic form - Darboux coordinates. More precisely,
we will use the Moser lemma \ct{moser} describing a cohomological
condition for two symplectic structures to be equivalent:
Let $M$ be a symplectic manifold of compact support.
Given two-forms $\omega$ and $\omega^\prime$ such that $[\omega]=[\omega^\prime] \in
H^2(M)$ and $\omega_t = \omega + t (\omega^\prime-\omega)$ is symplectic $\forall t
\in [0,1]$, then there exists a diffeomorphism $\phi: M \to M$ such that
$\phi^*(\omega^\prime)= \omega$. This implies that all $\omega_t$ are related
by coordinate transformations generated by a vector field $X$ satisfying
$\iota_X \omega_t + A=0$ where $\omega^\prime-\omega = dA$. In particular we
have $\phi^* (\omega^\prime) = \omega$ where $\phi$ is the flow of $X$.
In terms of local coordinates, there always exists a coordinate
transformation $\phi$ whose pullback maps $\omega^\prime = \omega +dA$
to $\omega$, i.e., $\phi: y \mapsto x= x(y)$ so that
\begin{equation}\label{darboux}
    \frac{\partial x^\alpha}{\partial y^\mu} \frac{\partial x^\beta}{\partial
    y^\nu} \omega^\prime_{\alpha\beta}(x) = \omega_{\mu\nu}(y).
\end{equation}

The Darboux theorem leads to an important consequence on the low
energy effective dynamics of D-branes in the presence of a
background $B$-field. The dynamics of D-branes is described by open
string field theory whose low energy effective action is obtained by
integrating out all the massive modes, keeping only massless fields
which are slowly varying at the string scale. The resulting
low energy dynamics is described by the Dirac-Born-Infeld (DBI) action \ct{dbi}.
For a $Dp$-brane in arbitrary background fields, the DBI action is given by
\begin{equation} \label{dbi-general}
S = \frac{2\pi}{(2\pi \kappa)^{\frac{p+1}{2}}}
\int d^{p+1} \sigma e^{-\Phi} \sqrt{\det(g + \kappa (B + F))}
+ {\cal O}(\sqrt{\kappa} \partial F, \cdots),
\end{equation}
where $\kappa \equiv 2 \pi \alpha^\prime$, the size of a string, is a
unique expansion parameter to control derivative corrections.
But the string coupling constant $g_s \equiv  e^{\langle \Phi \rangle}$ will be
assumed to be constant.

The DBI action \eq{dbi-general} respects several
symmetries.\footnote{The action \eq{dbi-general} has a worldvolume
reparameterization invariance: $\sigma^\mu \mapsto
{\sigma^\prime}^\mu = f^\mu(\sigma)$ for $\mu = 0,1, \cdots,p$.
But, this diffeomorphism symmetry is usually gauge-fixed
to identify worldvolume coordinates $\sigma^\mu$ with spacetime ones,
i.e., $x^A(\sigma) = \sigma^A$ for $A = 0,1,\cdots,p$.
In this static gauge which will be adopted throughout the paper,
the induced metric $g_{\mu\nu}(x(\sigma))$
on the brane reduces to a background spacetime metric,
e.g., $g_{\mu\nu}(\sigma) = \delta_{\mu\nu}$.
So we will never refer to this symmetry in our discussion.}
The most important symmetry for us is the so-called $\Lambda$-symmetry given by
\be \la{lambda-symmetry}
B \to B - d\Lambda, \quad A \to A + \Lambda
\ee
for any one-form $\Lambda$. Thus the DBI action depends
on $B$ and $F$ only in the gauge invariant combination $\CF \equiv B
+ F$ as shown in \eq{dbi-general}. Note that ordinary $U(1)$ gauge
symmetry is a special case where the gauge parameters $\Lambda$ are
exact, namely, $\Lambda = d \lambda$, so that $B \to B, \; A \to A +
d\lambda$.

Suppose that the two-form $B$ is closed, i.e. $dB=0$, and
non-degenerate on the D-brane worldvolume $M$. The pair $(M, B)$
then defines a symplectic manifold.\footnote{\la{general-geometry} Note that
the `D-manifold' $M$ also carries a non-degenerate, symmetric,
bilinear form $g$ which is a Riemannian metric. The pair $(M,g)$
thus defines a Riemannian manifold.
If we consider a general pair $(M, g+ \kappa B)$, it
describes a generalized geometry \ct{generalized-geometry} which continuously
interpolates between a symplectic geometry $(|\kappa Bg^{-1}| \gg 1)$ and
a Riemannian geometry $(|\kappa Bg^{-1}| \ll 1)$. The decoupling limit
considered in \ct{sw} corresponds to the former.} But the
$\Lambda$-transformation \eq{lambda-symmetry} changes (locally) the
symplectic structure from $\omega = B$ to $\omega^\prime = B
-d\Lambda$. According to the Darboux theorem and the Moser lemma
stated above, there must be a coordinate transformation such
as Eq.\eq{darboux}. Thus the local change of symplectic structure
due to the $\Lambda$-symmetry can always be translated into
worldvolume diffeomorphisms as in Eq.\eq{darboux}. For some reason
to be clarified later, we prefer to interpret the symmetry
\eq{lambda-symmetry} as a worldvolume diffeomorphism, denoted as $G
\equiv Diff(M)$, in the sense of Eq.\eq{darboux}. Note that the
number of gauge parameters in the $\Lambda$-symmetry is exactly the
same as $Diff(M)$. We will see that the Darboux theorem in symplectic
geometry plays the same role as the equivalence principle
in general relativity.

The coordinate transformation in Eq.\eq{darboux} is not unique since
the symplectic structure remains intact if it is generated by a
vector field $X$ satisfying ${\cal L}_X B = 0$. Since we are
interested in a simply connected manifold $M$, i.e. $\pi_1(M)=0$,
the condition is equivalent to $\iota_X B + d \lambda = 0$, in other
words, $X \in Ham(M)$.\footnote{If we consider NC gauge theories on
$M=\mathbf{T}^4$ in which $\pi_1(M) \neq 0$, a symplectic vector field $X$,
i.e. $\CL_X B = 0$, is not necessarily Hamiltonian
but takes the form $X^\mu = \theta^{\mu\nu} \p_\nu \lambda + \xi^\mu$
where $\xi^\mu$ is a harmonic one-form, i.e., it cannot be written as a derivative of a
scalar globally. This harmonic one-form introduces a twisting of vector bundle or
projective module on $M=\mathbf{T}^4$ such that
the gauge bundle is periodic only up to gauge transformations \ct{twisted-torus}.}
Thus the symplectomorphism $H \equiv Ham(M)$
corresponds to the $\Lambda$-symmetry where $\Lambda = d\lambda$ and
so $Ham(M)$ can be identified with the ordinary $U(1)$ gauge
symmetry \ct{cornalba,jurco-schupp}. As is well-known, if a vector field
$X_\l$ is Hamiltonian satisfying $\iota_{X_\l} B + d \lambda = 0$,
the action of $X_\l$ on a smooth function $f$ is given by
$X_\l(f) = \{\l, f\}$, which is infinite dimensional as well as non-Abelian
and, after quantization \eq{poisson-bracket},
gives rise to NC gauge symmetry.

Using the $\Lambda$-symmetry, gauge fields can always be
shifted to $B$ by choosing the parameters as $\Lambda_\mu = - A_\mu$,
and the dynamics of gauge fields in the new symplectic form $B + dA$
is interpreted as a local fluctuation of symplectic structures. This
fluctuating symplectic structure can then be translated into a
fluctuating geometry through the coordinate transformation in
$G=Diff(M)$ modulo $H=Ham(M)$, the $U(1)$ gauge transformation. We
thus see that the `physical' change of symplectic structures at a
point in $M$ takes values in
$Diff_F(M) \equiv G/H = Diff(M)/Ham(M)$.

We need an explanation about the meaning of the `physical'.
The $\Lambda$-symmetry \eq{lambda-symmetry} is spontaneously broken to
the symplectomorphism $H = Ham(M)$ since the vacuum manifold
defined by the NC spacetime \eq{nc-spacetime} picks up
a particular symplectic structure, i.e.,
\be \la{spacetime-vacuum}
\langle B_{\mu\nu}(x) \rangle_{\rm{vac}} = (\theta^{-1})_{\mu\nu}.
\ee
This should be the case since we expect only the ordinary $U(1)$ gauge
symmetry in large distance (commutative) regimes, corresponding
to $|\kappa Bg^{-1}| \ll 1$ in the footnote \ref{general-geometry}
where $|\theta|^2 \equiv G_{\mu\l} G_{\nu\s}
\theta^{\mu\nu} \theta^{\l\s} = \kappa^2 |\kappa Bg^{-1}|^2 \ll \kappa^2 $
with the open string metric $G_{\mu\nu}$ defined by Eq.(3.21) in \ct{sw}.
The fluctuation of gauge fields around the background \eq{spacetime-vacuum}
induces a deformation of the vacuum manifold, e.g. $\IR^4$ in the case of
constant $\theta$'s. According to the Goldstone's theorem \ct{goldstone},
massless particles, the so-called Goldstone bosons, should appear
which can be regarded as dynamical variables taking values
in the quotient space $G/H = Diff_F(M)$.

Since $G=Diff(M)$ is generated by the set of $\Lambda_\mu = - A_\mu$,
so the space of gauge field configurations on NC $\IR^4$ and $H=Ham(M)$
by the set of gauge transformations, $G/H$ can be identified with
the gauge orbit space of NC gauge fields, in other words, the `physical'
configuration space of NC gauge theory. Thus the moduli space of all possible
symplectic structures is equivalent to the `physical' configuration
space of NC electromagnetism.

The Goldstone bosons for the spontaneous symmetry breaking $G \to H$ turn out
to be spin-2 gravitons \ct{g/h}, which might be elaborated by the following
argument. Using the coordinate transformation \eq{darboux} where $\w^\prime = B+F(x)$
and $\w = B$, one can get the following identity \cite{cornalba}
for the DBI action \eq{dbi-general}
\begin{equation} \label{proof-sw}
\int d^{p+1} x \sqrt{\det(g + \kappa (B + F(x)))}
= \int d^{p+1} y \sqrt{\det(\kappa B + h(y))},
\end{equation}
where fluctuations of gauge fields now appear as an induced metric on
the brane given by
\begin{equation} \label{induced-metric}
h_{\mu\nu}(y) =  \frac{\partial x^\alpha}{\partial y^\mu}
\frac{\partial x^\beta}{\partial y^\nu} g_{\alpha\beta}.
\end{equation}
The dynamics of gauge fields is then encoded into the fluctuations
of geometry through the embedding functions $x^\mu(y)$.
The fluctuation of gauge fields around the background \eq{spacetime-vacuum}
can be manifest by representing the embedding function as follows
\be \la{cov-coord}
x^\mu(y) \equiv y^\mu + \theta^{\mu\nu} {\widehat A}_\nu(y).
\ee
Given a gauge transformation $A \to A + d \l$, the corresponding coordinate
transformation generated by a vector field $X_\l \in Ham(M)$ is given by
\bea \la{nc-gauge-tr}
\d x^\mu(y) &\equiv& X_\l (x^\mu) =
- \{\l, x^\mu \} \xx &=& \theta^{\mu\nu}(\p_\nu \l + \{ {\widehat A}_\nu, \l \}).
\eea
As we discussed already, this coordinate change can be identified
with a NC gauge transformation after the quantization \eq{poisson-bracket}.
So $ {\widehat A}_\mu(y)$ are NC $U(1)$ gauge fields and the coordinates
$x^\mu(y)$ in \eq{cov-coord} will play a special role since they are gauge
covariant \ct{madore} as well as background independent \ct{seiberg}.

It is straightforward to get the relation between ordinary and NC
field strengths from the identity \eq{darboux}:
\be \la{semi-esw}
\left(\frac{1}{B+F(x)}\right)^{\mu\nu} = \Bigl(\theta - \theta
\what{F}(y) \theta \Bigr)^{\mu\nu} \;\; \Leftrightarrow \;\;
\what{F}_{\mu\nu}(y) = \left(\frac{1}{1+F\theta} F
\right)_{\mu\nu}(x),
\ee
where NC electromagnetic fields are defined by
\be \la{semi-nc-f}
\widehat{F}_{\mu\nu}=\partial_{\mu}\widehat{A}_{\nu} -
\partial_{\nu}\widehat{A}_{\mu} + \{\widehat{A}_{\mu},
\widehat{A}_{\nu} \}.
\ee
The Jacobian of the coordinate transformation $y \mapsto x= x(y)$ is obtained
by taking the determinant on both sides of Eq.\eq{darboux}
\be \label{sw-measure}
d^{p+1} y = d^{p+1} x \sqrt{\det(1+ F \theta)}(x).
\ee
In addition one can show \cite{cornalba} that the DBI action (\ref{proof-sw})
turns into the NC gauge theory with the semi-classical field
strength \eq{semi-nc-f} by expanding the right-hand side with
respect to $h/\kappa B$ around the background $B$.

The above argument clarifies why the dynamics of NC gauge fields can be
interpreted as the fluctuations of geometry described by
the metric (\ref{induced-metric}). One may identify
$\partial x^\alpha/\partial y^\mu \equiv e^\a_\mu (y)$ with vielbeins on some
manifold $\CM$ by regarding $h_{\mu\nu}(y)= e^\a_\mu(y) e^\b_\nu(y) g_{\a\b}$
as a Riemannian metric on $\CM$.
The embedding functions $x^\mu(y)$ in \eq{cov-coord}, which are now dynamical
fields, subject to the equivalence relation, $x^\mu \sim x^\mu + \d x^\mu$,
defined by the gauge transformation \eq{nc-gauge-tr}, coordinatize the
quotient space $G/H = Diff_F(M)$. As usual, $y^\mu$ are vacuum expectation
values of $x^\mu$ specifying the background \eq{spacetime-vacuum} and
$\what{A}_\mu(y)$ are fluctuating (dynamical) coordinates (fields).
In this context, the gravitational fields $e^\a_\mu (y)$ or $h_{\mu\nu}(y)$
correspond to the Goldstone bosons for the spontaneous symmetry breaking
\eq{spacetime-vacuum}. This is a rough picture showing how gravity
can emerge from NC electromagnetism.

So far we are mostly confined to semi-classical limit, say $\CO(\hbar)$
in Eq.\eq{star-product}. The semi-classical means here
slowly varying fields, $\sqrt{\kappa} |\frac{\partial F}{F}| \ll 1$,
in the sense keeping field strengths (without restriction on their
size) but not their derivatives. We will consider derivative corrections
in the coming sections. This paper is organized as follows.

In section 2, we will revisit the equivalence between ordinary
and NC DBI actions shown by Seiberg and Witten \ct{sw}.
We will show that the exact Seiberg-Witten (SW) map in Eqs.\eq{semi-esw} and
\eq{sw-measure} are a direct consequence of the equivalence
after a simple change of variables between open and closed strings
as was shown in \ct{hsy,ban-yang}. This argument illuminates why
higher order terms in Eq.\eq{star-product} correspond to derivative corrections
$\CO(\sqrt{\kappa} \partial F)$ in the DBI action \eq{dbi-general}.
The leading four-derivative corrections were completely
determined by Wyllard \ct{wyllard1}.
We will argue that the SW map with derivative corrections should be
obtained from the Wyllard's result by the same change of variables
between open and closed string parameters.
Since our main goal in this paper is to elucidate the relation between NC
gauge theory and gravity, we will not explicitly check the identities so
naturally emerging from well-established relations. Rather they could be
regarded as our predictions. According to the correspondence between NC gauge
theory and gravity, it is natural to expect that the derivative
corrections give rise to higher order gravity, e.g., $R^2$ gravity.

In section 3, we will newly derive the SW map for the derivative corrections
in the context of deformation quantization \ct{d-quantize,kontsevich}.
The deformation quantization provides a noble approach to reify the
Darboux theorem beyond the semiclassical, i.e. $\CO(\theta)$, limit. For
example, the SW maps, Eq.\eq{semi-esw} and Eq.\eq{sw-measure}, result from the
equivalence in the $\CO(\theta)$ approximation between
the star products $\star_{\omega^\prime}$ and $\star_\omega$ defined
by the symplectic forms  $\omega^\prime = B + F(x)$
and $\omega=B$, respectively \ct{jurco-schupp,liu}.
In a seminal paper, M. Kontsevich proved \ct{kontsevich} that every
finite-dimensional Poisson manifold $M$ admits a canonical deformation
quantization. Furthermore he proved that, changing coordinates in a star product,
one obtains another gauge equivalent star product. This was explicitly checked
in \ct{zotov} by making an arbitrary change of coordinates, $y^\mu \mapsto
x^\mu(y)$, in the Moyal $\star$-product \eq{star-product} and obtaining
the deformation quantization formula up to the third order. This result is
consistent with the SW map in section 2 about derivative corrections.
After inspecting the basic principle of deformation quantization, we
put forward a conjecture that the emergent gravity from NC
electromagnetism corresponds to a nonlinear realization $G/H$ of the
diffeomorphism group or more generally its NC deformation,
so meeting a framework of NC gravity \ct{nc-gravity,nc-gravity2}.
(See also a review \ct{szabo}.)

In section 4, we will explore the equivalence between NC $U(1)$ instantons and
gravitational instantons found in \ct{sty,ys} to illustrate the
correspondence of NC gauge theory with gravity. The emergent gravity
reveals a remarkable feature that self-dual NC
electromagnetism nicely fits with the twistor space describing
curved self-dual spacetime \ct{penrose,atiyah}.
This construction, which closely follows the results on $N=2$
strings \ct{ooguri-vafa-mpl,ooguri-vafa}, will also clarify how the deformation of
symplectic (or K\"ahler) structure on $\IR^4$ due to the fluctuation of gauge fields
appears as that of complex structure of the twistor space \ct{mine}.
We observe that our construction is remarkably in parallel with
topological D-branes on NC manifolds \ct{kapustin}, suggesting a possible
connection with the generalized complex geometry \ct{generalized-geometry}.

In section 5, we will generalize the equivalence in section 4 using the
background independent formulation of NC gauge theories \ct{sw,seiberg}
and show that self-dual electromagnetism in NC spacetime
is equivalent to self-dual Einstein gravity \ct{mine}.
This section will also serve to uncover many details in \ct{mine}.
In the course of the construction, it becomes obvious that
a framework of NC gravity is in general needed in order to incorporate
the full quantum deformation of diffeomorphism symmetry.
We will also discuss in detail the twistor space structure
inherent in the self-dual NC electromagnetism.

Finally, in section 6, we will raise several open issues in the emergent
gravity from NC spacetime and speculate possible implications
for the correspondence between NC gauge theory and gravity.

\section{Derivative Corrections and Exact SW Map}

We revisit here the equivalence between NC and ordinary gauge theories
discussed in \ct{sw}. First let us briefly recall how NC gauge theory arises in
string theory. The coupling of an open string attached on a D$p$-brane to
massless backgrounds is described by a sigma model of the form
\be \la{string-action}
S = \frac{1}{2\kappa}\int_{\Sigma} d^2 \sigma (g_{\mu\nu} (x) \p_a
x^\mu \p^a x^\nu - i \kappa \varepsilon^{ab} B_{\mu\nu} (x)  \p_a
x^\mu \p_b x^\nu ) - i \int_{\p \Sigma} d\tau A_\mu (x) \p_\tau x^\mu,
\ee
where string worldsheet $\Sigma$ is the upper half plane parameterized by $-\infty
\leq \tau \leq \infty$ and $0 \leq \sigma \leq \pi$ and $\p \Sigma$ is its boundary.
The $\Lambda$-symmetry \eq{lambda-symmetry}, which underlies
the emergent gravity, is obvious by rewriting relevant terms into
form language, $ \int_{\Sigma} B  + \int_{\p \Sigma} A$, as a simple application of
Stokes' theorem.

We leave the geometry of closed string backgrounds fixed and
concentrate, instead, on the dynamics of open string sector. To be
specific, we consider flat spacetime with the constant Neveu-Schwarz
$B$-field. Here we regard the last term in Eq.\eq{string-action} as an
interaction with background gauge fields and define the propagator
in terms of free fields $y^\mu(\tau,\s)$ subject to the boundary conditions
\be \la{boundary-condition}
g_{\mu\nu} \p_\s y^\nu + i \kappa B_{\mu\nu} \p_\tau y^\nu|_{\p \Sigma}
= 0,
\ee
where the worldsheet fields $x^\mu(\tau,\s)$ were simply renamed $y^\mu(\tau,\s)$
to compare them with another free fields satisfying different boundary
conditions, e.g., $g_{\mu\nu} \p_\s x^\nu |_{\p \Sigma} = 0$.
The propagator evaluated at boundary points \ct{sw} is
\begin{equation}\label{open-propagator}
    \langle y^\mu (\tau)  y^\nu (\tau^\prime) \rangle = -
    \frac{\kappa}{2\pi} \Bigl(\frac{1}{G}\Bigr)^{\mu\nu} \log(\tau - \tau^\prime)^2 +
    \frac{i}{2}  \theta^{\mu\nu} \epsilon(\tau - \tau^\prime)
\end{equation}
where $\epsilon(\tau)$ is the step function. Here
\bea \la{open-g}
&& G_{\mu\nu} = g_{\mu\nu} - \kappa^2 (B g^{-1}B)_{\mu\nu}, \\
\la{open-g-inverse}
&& \left(\frac{1}{G}\right)^{\mu\nu} = \left( \frac{1}{g + \kappa B}
g \frac{1}{g - \kappa B}\right)^{\mu\nu}, \\
\la{open-theta}
&& \theta^{\mu\nu} = - \kappa^2 \left( \frac{1}{g + \kappa B}B
\frac{1}{g - \kappa B}\right)^{\mu\nu}.
\eea
They are related via the following relation
\begin{equation} \label{op-cl}
\frac{1}{G} + \frac{\theta}{\kappa} =  \frac{1}{g + \kappa B}.
\ee
The metric $G_{\mu\nu}$ has a simple interpretation as an
effective metric seen by open strings while $g_{\mu\nu}$ is
the closed string metric. Furthermore the parameter
$\theta^{\mu\nu}$ can be interpreted as the noncommutativity in a space where
embedding coordinates on a D$p$-brane describe the NC
coordinates \eq{nc-spacetime}.

For a moment we will work in the approximation of slowly varying
fields relative to the string scale, in the sense of neglecting
derivative terms, i.e., $ \sqrt{\kappa} |\frac{\p F}{F}| \ll 1$,
but of no restriction on the size of field strengths.
Nevertheless, the field strengths $F$ need not be constant.
Indeed the field strength can vary rapidly in the sense of low energy field
theory as long as a typical length scale of the varying $F$ is much larger
than the string scale. In this limit the open string effective action
on a D-brane is given by the DBI action \ct{dbi}.
Seiberg and Witten, however, showed \ct{sw} that an explicit form of the
effective action depends on the regularization scheme of two dimensional field
theory defined by the worldsheet action \eq{string-action}.
The difference due to different regularizations is always in a choice of
contact terms, leading to the redefinition of coupling constants which are
spacetime fields. So theories defined with different regularizations
are related each other by the field redefinitions in spacetime.

The usual infinities in quantum field theory also arise in the worldsheet path
integral defined by the action \eq{string-action} and the theory has to be
regularized. Using the propagator \eq{open-propagator} with a point-splitting
regularization \ct{sw} where different operators are never at the same
point, the spacetime effective action is expressed in terms of NC gauge fields
and has the NC gauge symmetry on the NC spacetime \eq{nc-spacetime}.
In this description, the analog of Eq.\eq{dbi-general} is
\begin{equation}\label{dbi-nc}
\widehat{S}(G_s, G, \widehat{A}, \theta) =
\frac{2\pi}{G_s (2\pi \kappa)^{\frac{p+1}{2}}} \int d^{p+1} y
\sqrt{\det(G + \kappa \widehat{F})} + \CO(\sqrt{\kappa}
\widehat{D} \widehat{F}).
\end{equation}
The action depends only on the open string variables $G_{\mu\nu}, \theta_{\mu\nu}$
and $G_s$, where the $\theta$-dependence is entirely in the
$\star$-product in the NC field strength
\be \la{nc-f}
\widehat{F}_{\mu\nu} = \p_\mu \widehat{A}_\nu -  \p_\nu \widehat{A}_\mu
- i [\widehat{A}_\mu, \widehat{A}_\nu]_\star.
\ee
The DBI action \eq{dbi-nc} is invariant under
\begin{equation}\label{nc-gt}
\widehat{\delta}_{\widehat{\lambda}} \widehat{A}_\mu =
\widehat{D}_\mu \widehat{\lambda} = \p_\mu \widehat{\lambda} - i
[\widehat{A}_\mu, \widehat{\lambda}]_\star.
\end{equation}
The NC field strength \eq{nc-f} and the NC gauge transformation \eq{nc-gt}
are the quantum version of  Eq.\eq{semi-nc-f} and Eq.\eq{nc-gauge-tr},
respectively, in the sense of Eq.\eq{poisson-bracket}.

Since the sigma model \eq{string-action} respects
the $\Lambda$-symmetry \eq{lambda-symmetry},
one can absorb the constant $B$-field
completely into gauge fields by choosing the gauge parameter
$\Lambda_\mu = -\half B_{\mu\nu} x^\nu$. The worldsheet action
is then given by
\be \la{string-action-2}
S = \frac{1}{2\kappa}\int_{\Sigma} d^2 \sigma g_{\mu\nu} \p_a
x^\mu \p^a x^\nu - i \int_{\p \Sigma} d\tau \Bigl(A_\mu (x) -
\half B_{\mu\nu} x^\nu \Bigr) \p_\tau x^\mu.
\ee
Now we regard the second part as the boundary interaction and define
the propagator with the first part with the boundary condition
$g_{\mu\nu} \p_\s x^\nu |_{\p \Sigma} = 0$, resulting in the usual Neumann
propagator.

The sigma model path integral using the Neumann propagator with Pauli-Villars
regularization, for example, preserves the ordinary gauge symmetry of
open string gauge fields \ct{sw}. In this case, the spacetime low energy effective
action on a single D$p$-brane, which is denoted as $S(g_s, g, A, B)$ to
emphasize the background dependence, is given by the DBI action \eq{dbi-general}.
Note that the effective action is now expressed in terms of closed
string variables $g_{\mu\nu}, B_{\mu\nu}$ and $g_s$.

Since the commutative and NC descriptions arise from the same open
string theory depending on different regularizations and the physics
should not depend on the regularization scheme, one may expect that
\begin{equation}\label{equiv-dbi}
    \widehat{S}(G_s, G, \widehat{A}, \theta) = S(g_s, g, A, B) +
    {\cal O}(\sqrt{\kappa} \partial F).
\end{equation}
If so, the two descriptions should be related each other by a spacetime
field redefinition. Indeed, Seiberg and Witten showed the identity
\eq{equiv-dbi} and also found the transformation, the so-called SW map,
between ordinary and NC gauge fields in such a way that preserves the
gauge equivalence relation of ordinary and NC gauge symmetries \ct{sw}.
The equivalence \eq{equiv-dbi} can also be understood as
resulting from different path integral prescriptions \ct{tlee,andreev-dorn}
based on the $\Lambda$-symmetry as we discussed above.
First of all, the equivalence \eq{equiv-dbi} determines the relation between
open and closed string coupling constants from the fact that for $F=\widehat{F}=0$
the constant terms in the actions using the two set of variables
are the same:
\begin{equation}\label{Gs-gs}
G_s = g_s \sqrt{\frac{\det G}{\det (g + \kappa B)}}.
\end{equation}

As was explained in \ct{sw}, there is a general description with an arbitrary $\theta$
associated with a suitable regularization that interpolates
between Pauli-Villars and point-splitting. This freedom is
basically coming from the $\Lambda$-symmetry that imposes the gauge invariant
combination of $B$ and $F$ in the open string theory as ${\cal F} = B+F$.
Thus there is a symmetry of shift in $B$ keeping $B+F$ fixed. Given
such a symmetry, we may split the $B$-field into two parts and put
one in the kinetic part and the
rest in the boundary interaction part. By taking the background to be $B$ or $B^\prime$,
we should get a NC description with appropriate $\theta$ or
$\theta^\prime$, and different $\widehat{F}$'s.
Hence we can write down a differential equation that describes how
$\widehat{A}(\theta)$ and $\widehat{F}(\theta)$ should change, when
$\theta$ is varied, to describe equivalent physics \ct{sw}:
\begin{eqnarray}\label{sw-de-a}
 &&   \delta\widehat{A}_\mu (\theta) =
    -\frac{1}{4}\delta\theta^{\alpha\beta} \Bigl( \widehat{A}_\alpha \star
    (\partial_\beta \widehat{A}_\mu + \widehat{F}_{\beta\mu} )
+ (\partial_\beta \widehat{A}_\mu + \widehat{F}_{\beta\mu}) \star
\widehat{A}_\alpha \Bigr), \\
\label{sw-de-f}
&&   \delta\widehat{F}_{\mu\nu} (\theta) =
    \frac{1}{4}\delta\theta^{\alpha\beta} \Bigl( 2\widehat{F}_{\mu\alpha} \star
    \widehat{F}_{\nu\beta} + 2\widehat{F}_{\nu\beta} \star
    \widehat{F}_{\mu\alpha} -
    \widehat{A}_\alpha \star (\widehat{D}_\beta \widehat{F}_{\mu \nu}
    + \p_\beta \widehat{F}_{\mu\nu}) \nonumber \\
&& \hspace{2.7cm} - (\widehat{D}_\beta \widehat{F}_{\mu \nu}
    + \p_\beta \widehat{F}_{\mu\nu}) \star \widehat{A}_\alpha
    \Bigr).
\end{eqnarray}
An exact solution of the differential equation \eq{sw-de-f} in the Abelian case
was found in \ct{liu,exact-sw-map}.

The freedom in the description just explained above is parameterized by a
two-form $\Phi$ from the viewpoint of NC geometry on D-brane worldvolume.
The change of variables for the general case is given by
\bea \la{op-cl-gen}
&& \frac{1}{G + \kappa \Phi} + \frac{\theta}{\kappa}
=  \frac{1}{g + \kappa B}, \\
\label{Gs-gs-gen}
&& G_s = g_s \sqrt{\frac{\det (G + \kappa \Phi)}{\det (g + \kappa B)}}.
\eea
The effective action with these variables are modified to
\begin{equation}\label{ncdbi-gen}
\widehat{S}_\Phi(G_s, G, \widehat{A}, \theta) =
\frac{2\pi}{G_s (2\pi \kappa)^{\frac{p+1}{2}}} \int d^{p+1} y
\sqrt{\det(G + \kappa (\widehat{F}+ \Phi))}.
\end{equation}
For every background characterized by $B, g_{\mu\nu}$ and $g_s$,
we thus have a continuum of descriptions labeled by a choice of
$\Phi$. So we end up with the most general form of the equivalence
for slowly varying fields, i.e.,$ \sqrt{\kappa} |\frac{\p F}{F}| \ll 1$:
\begin{equation}\label{equiv-dbi-gen}
 \widehat{S}_\Phi(G_s, G, \widehat{A}, \theta)
= S(g_s, g, A, B) +   {\cal O}(\sqrt{\kappa} \partial F),
\end{equation}
which was proved in \ct{sw} using the
differential equation \eq{sw-de-f} and the change
of variables, \eq{op-cl-gen} and \eq{Gs-gs-gen}.

The above change of variables between open and closed string parameters
is independent of dynamical gauge fields
and so one can freely use them independently of local
dynamics to express two different descriptions with the same string variables \ct{hsy}.
For example, we get from Eq.\eq{op-cl-gen}
\begin{equation}\label{sw-metric}
\CG \equiv g + \kappa (B + F) =  (1+ F \theta) \Bigl( G + \kappa
(\Phi + {\mathbf F}) \Bigr) \frac{1}{1 + \frac{\theta}{\kappa}(G+\kappa\Phi)}
\end{equation}
where
\be \la{bold-f}
{\mathbf F} (x) = \left(\frac{1}{1+F\theta} F\right)(x).
\ee
The equivalence \eq{equiv-dbi-gen}, using the identity \eq{sw-metric},
immediately leads to the dual description of the NC DBI action via the
exact SW map \ct{hsy,ban-yang}
\begin{eqnarray}\label{sw-equiv}
 &&  \int d^{p+1} y \sqrt{\det(G + \kappa ( \Phi + \widehat{F}))} \nonumber \\
 && \hspace{2cm} = \int d^{p+1} x \sqrt{\det{(1+ F \theta})}
 \sqrt{\det{(G + \kappa (\Phi + {\bf F}))}}
+  {\cal O}(\sqrt{\kappa}\partial F).
\end{eqnarray}
Note that the commutative action in Eq.\eq{sw-equiv} is exactly the same as
the DBI action obtained from the worldsheet sigma model
using $\zeta$-function regularization scheme \ct{andreev-dorn}.
The equivalence \eq{sw-equiv} was also proved in \ct{jusch-wess}
in the framework of deformation quantization.

One can expand both sides of Eq.\eq{sw-equiv} in powers of $\kappa$.
${\cal O}(1)$ implies that there is a measure change
between NC and commutative descriptions,
which is exactly the same as Eq.\eq{sw-measure}.
In other words, the coordinate transformation, $y^\mu \mapsto x^\mu(y)$,
between commutative and NC descriptions depends on the dynamical gauge fields.
Since the identity \eq{sw-equiv} must be true for arbitrary small $\kappa$,
substituting Eq.\eq{sw-measure} into Eq.\eq{sw-equiv} leads to the
relation \eq{semi-esw}, but now for the NC field strength \eq{nc-f}.
We thus see that the embedding coordinates $x^\mu(y)$
are always defined by \eq{cov-coord} independently of the choice
$\Phi$. This is consistent with the fact \ct{seiberg} that the covariant
coordinates $x^\mu(y)$ are background independent.

So far we have ignored derivative terms containing $\p^n F$. However the
left hand side of Eq.\eq{sw-equiv} contains infinitely many derivatives from
the star commutator in $\what{F}$ which have to generate such derivative terms,
though we had taken the ordinary product neglecting a potential NC
ordering. So we need to carefully look into the identity
\eq{sw-equiv} to what extent the equivalence holds. In fact it is
easily inferred from the SW map in \ct{fidanza} that the
left hand side of Eq.\eq{sw-equiv} contains infinitely many
higher order derivative terms. The derivative corrections are coming
from $F_{\mu\nu}^{(n,m)}$ for $n < m$ with the notation in \ct{fidanza}
(see the figure 1 and section 3.2). This can also be inferred from the
previous argument related to the SW map \eq{semi-esw} which does not
incorporate any derivative corrections and precisely corresponds to
$F_{\mu\nu}^{(n,n)}$ in \ct{fidanza}. As we discussed there,
Eq.\eq{semi-esw} is the SW map for the semi-classical
field strength \eq{semi-nc-f} and the DBI action \eq{proof-sw}
is equivalent to the semi-classical DBI action \ct{cornalba}
where field strengths are given by \eq{semi-nc-f} rather than \eq{nc-f}.
It is thus obvious that the equivalence \eq{sw-equiv} is still true
with the field strength \eq{semi-nc-f} in the approximation
of slowly varying fields.

So there must be more terms with derivative corrections on the right
hand side of Eq.\eq{sw-equiv} if one insists to keep the NC field strength \eq{nc-f}.
To find the derivative corrections systematically, however, one has
to notice that there are another sources giving rise to them.
The NC description has two dimensionful parameters, $\theta$ and $\kappa$,
which control derivative corrections. The parameter $\kappa$ takes into
account stringy effects coming from massive modes in worldsheet conformal
field theory, as indicated in Eq.\eq{dbi-nc},
while the noncommutativity parameter $\theta$ does the effects of
NC spacetime in worldvolume field theory. Which one becomes more important
depends on a scale we are probing.

We are interested in the Seiberg-Witten limit \ct{sw}, $\kappa \to 0$,
keeping all open string variables fixed. In this limit, $|\theta|^2 \equiv
G_{\mu\l} G_{\nu\s} \theta^{\mu\nu} \theta^{\l\s} = \kappa^2 |\kappa
Bg^{-1}|^2 \gg \kappa^2$, using the metric $G_{\mu\nu}$ in
the background independent scheme, i.e., $\Phi = -B$ in Eq.\eq{op-cl-gen}.
This implies that the noncommutativity effects in the SW limit are
predominant compared to stringy effects.
So we will neglect the stringy effects such as $\CO(\sqrt{\kappa}
\widehat{D} \widehat{F})$ in Eq.\eq{dbi-nc} \ct{derivative-dms}.
But we have to keep $\kappa \what{F}$ since $\what{F}$ could be arbitrarily large.
The stringy corrections in NC
gauge theory have been discussed in several papers \ct{dbi-derivative}
based on the SW equivalence between ordinary and NC gauge theories.

An ordering problem in NC spacetime has to be taken into account.
A unique feature is that translations in NC directions
are basically gauge transformations, i.e.,
$e^{ik \cdot y} \star f(y) \star e^{-ik \cdot y} = f(y + k \cdot \theta)$.
This immediately implies that there are no local gauge-invariant
observables in NC gauge theory \cite{non-local}.
It turns out that NC gauge theories allow a new type of gauge invariant
objects, the so-called open Wilson lines, which are localized in momentum
space \ct{open-wilson}. Attaching local operators which transform in the
adjoint representation of gauge transformations to an open Wilson line
also yields gauge invariant operators \ct{non-local}.
For example, the NC DBI action carrying a definite momentum $k$ is given by
\begin{equation}\label{nc-dbi-k}
    \widehat{S}_\Phi^k (G_s, G, \widehat{A}, \theta)
    = \frac{2\pi}{G_s (2\pi \kappa)^{\frac{p+1}{2}}}
    \int d^{p+1} y L_\star \Bigl[ \sqrt{\det(G + \kappa (\widehat{F} +
    \Phi))} W(y, C_k) \Bigr] \star e^{i k \cdot y},
\end{equation}
where $W(y, C_k)$ is a straight Wilson line with momentum $k$ with
path $C_k$ and $L_\star$ is defined as smearing the operators
along the Wilson line and taking the path ordering with respect to
$\star$-product. We refer \ct{liu} for more informations useful for
Eq.\eq{nc-dbi-k}. The DBI action \eq{ncdbi-gen} corresponds to
$\widehat{S}_\Phi^{k=0}$ without regard to the NC ordering.

Let us now turn to the commutative description. Unlike the NC case, there is only
one dimensionful parameter, $\kappa$, to control derivative corrections.
So the derivative corrections due to $\theta$ and $\kappa$ in NC gauge theory
all appear as stringy corrections from the viewpoint of commutative
description and they are intricately entangled. The derivative correction
to the DBI action \eq{dbi-general} has been calculated by Wyllard \ct{wyllard1}
using the boundary state formalism and it is given by
\bea \label{der-correction-c}
&& S_W(g_s, g, A, B) = \frac{2\pi}{g_s (2\pi \kappa)^{\frac{p+1}{2}}}
    \int d^{p+1} x \sqrt{\det \CG}\Bigl( 1 + \frac{\kappa^4}{96}
    (-\CG^{\mu_4 \mu_1}\CG^{\mu_2 \mu_3}\CG^{\rho_4 \rho_1}\CG^{\rho_2 \rho_3}
    S_{\rho_1\rho_2\mu_1\mu_2}S_{\rho_3\rho_4\mu_3\mu_4} \xx
&& \hspace{3cm}    + \half \CG^{\rho_4 \rho_1}\CG^{\rho_2 \rho_3}
    S_{\rho_1\rho_2}S_{\rho_3\rho_4}) + \cdots \Bigr),
\eea
where $\CG_{\mu\nu}$ is a non-symmetric metric defined by \eq{sw-metric}
and $\CG^{\mu\nu}$ is its inverse, i.e., $\CG^{\mu\lambda} \CG_{\lambda\nu} =
{\delta^\mu}_\nu$. The tensor
\begin{equation}\label{curvature-s}
    S_{\rho_1\rho_2\mu_1\mu_2} = \p_{\rho_1}\p_{\rho_2}
    F_{\mu_1\mu_2} + \kappa \CG^{\nu_1\nu_2} \Bigl( \p_{\rho_1} F_{\mu_1\nu_1}
    \p_{\rho_2} F_{\mu_2\nu_2} -  \p_{\rho_1} F_{\mu_2\nu_1}
    \p_{\rho_2} F_{\mu_1\nu_2} \Bigl)
\end{equation}
may be interpreted as the Riemann tensor for the nonsymmetric
metric $\CG_{\mu\nu}$ and $S_{\rho_1\rho_2} = \CG^{\mu_1 \mu_2}
S_{\rho_1\rho_2\mu_1\mu_2}$ as the Ricci tensor \ct{wyllard1}.

As we reasoned before, the SW equivalence between ordinary and NC
gauge theories has to be general regardless of a specific limit
under our consideration. So this should be the case even after
including derivative corrections in ordinary and NC theories. Let us
denote these corrections by
$\Delta S_{DBI}$ and $\Delta \what{S}_{DBI}$, respectively.
The SW equivalence in general means that
\be \la{general-sw-equivalence}
S_{DBI} + \Delta S_{DBI} = \what{S}_{DBI} + \Delta \what{S}_{DBI}.
\ee
We already argued that we will neglect the NC correction $\Delta
\what{S}_{DBI}$ in the SW limit. We will discuss later to what extent we
can do it. The equivalence \eq{general-sw-equivalence} in this limit
then reduces to \ct{derivative-dms}
\be \la{sw-sw-equivalence}
S_{DBI}|_{SW} + \Delta S_{DBI}|_{SW}
= \what{S}_{DBI}|_{SW}.
\ee

Recall that the exact SW map \eq{sw-equiv} was obtained by the equivalence
\eq{equiv-dbi-gen} with the simple change
of variables between open and closed string parameters
defined by Eqs.\eq{op-cl-gen} and \eq{Gs-gs-gen}.
This change of variables should be true even with derivative corrections
since they are independent of local dynamics.
As illustrated in Eq.\eq{sw-equiv}, the description of both DBI actions in terms
of the same string variables has provided a great simplification to identify SW maps.
Thus we will equally use the open string variables for the derivative corrections in
Eq.\eq{der-correction-c}, where the metric $\CG_{\mu\nu}$ will be replaced by
Eq.\eq{sw-metric}. Since we are commonly using the open string variables for both
descriptions, the SW limit in \eq{sw-sw-equivalence} simply means the zero
slope limit, i.e. $\kappa \to 0$. So it is straightforward to extract the SW
maps with derivative corrections by expanding both sides of
Eq.\eq{sw-sw-equivalence} in powers of $\kappa$.

Though the general case with $\Phi$ does not introduce any complication, we
will work in the background independent scheme where $\Phi = -B = -
1/\theta$, for definiteness. In this case, the metric $\CG_{\mu\nu}$ has a
simple expression
\begin{equation}\label{sw-metric-bi}
   \CG_{\mu\nu} = \kappa \Bigl({\rm g} \theta^{-1} -
\frac{\kappa}{\theta G \theta} \Bigr)_{\mu\nu},
\qquad \CG^{\mu\nu} = \frac{1}{\kappa} \left( \theta {\rm g}^{-1}
\Bigl( 1 - \frac{\kappa}{{\rm g} G \theta} \Bigr)^{-1} \right)^{\mu\nu},
\end{equation}
where we have introduced an effective metric ${\rm g}_{\mu\nu}$ induced by
dynamical gauge fields
\begin{equation}\label{effective-metric}
    {\rm g}_{\mu\nu} = \delta_{\mu\nu} + (F\theta)_{\mu\nu},
    \qquad  ({\rm g}^{-1})^{\mu\nu} \equiv {\rm g}^{\mu\nu} =
    \Bigl(\frac{1}{1 + F\theta}\Bigr)^{\mu\nu},
\end{equation}
which will play a role in our later discussions. (Unfortunately we
have abused many metrics, $(g,G,h,\CG,{\rm g})$. We hope it does not cause
many confusions in discriminating them.) Noting that
${\rm g} \theta^{-1} = B+ F$, Eq.\eq{darboux} implies that the
effective metric ${\rm g}_{\mu\nu}$ is not independent of the induced metric
$h_{\mu\nu}$ in Eq.\eq{induced-metric} but related as follows:
\be \la{metric-h-g}
h_{\mu\nu}(y)= e^\a_\mu(y) e^\b_\nu(y) g_{\a\b},
\qquad (\theta {\rm g}^{-1})^{\a\b}(y) = \theta^{\mu\nu} e^\a_\mu(y) e^\b_\nu(y).
\ee
Identifying $ e^\a_\mu (y)\equiv \partial x^\alpha/\partial y^\mu$
with vielbeins on some emergent manifold $\CM$, it is suggestive
that the Darboux theorem can be interpreted as the equivalence
principle in symplectic geometry.

Let us start with $\CO(1)$ terms from both sides in Eq.\eq{sw-sw-equivalence}.
Note that the factor $\kappa^4$ in front of the derivative corrections
in Eq.\eq{der-correction-c} is precisely cancelled by the factors from the metric
$\CG^{-1}$ in Eq.\eq{sw-metric-bi}, and thus they already give rise to the $\CO(1)$
contribution. Taking this into account, we get the following SW map
\footnote{Although we use the momentum space representation
for the manifest gauge invariance, the actual comparison with the
commutative description is understood to be made in coordinate space
using the formula (2.16) in \ct{liu}.}
\bea \la{sw-derivative-0}
&& \int d^{p+1} y L_\star \Bigl[ W(y, C_k) \Bigr]
\star e^{i k \cdot y} \xx
&=& \int d^{p+1} x \sqrt{\det {\rm g}} \Bigl( 1 - \frac{1}{96}
{\rm g}^{\mu_4 \mu_1}{\rm g}^{\mu_2 \mu_3}(\theta {\rm g}^{-1})^{\rho_4 \rho_1}
(\theta {\rm g}^{-1})^{\rho_2 \rho_3} \mathbf{S}_{\rho_1\rho_2\mu_1\mu_2}
\mathbf{S}_{\rho_3\rho_4\mu_3\mu_4} \xx
&& \hspace{3cm} + \frac{1}{192} (\theta {\rm g}^{-1})^{\rho_4 \rho_1}
(\theta {\rm g}^{-1})^{\rho_2 \rho_3} \mathbf{S}_{\rho_1\rho_2}
\mathbf{S}_{\rho_3\rho_4} + \cdots \Bigr),
\eea
where
\begin{equation}\label{eff-curvature}
\mathbf{S}_{\rho_1\rho_2\mu_1\mu_2} = \p_{\rho_1}\p_{\rho_2}
    {\rm g} _{\mu_1\mu_2} -  {\rm g}^{\nu_1\nu_2} \Bigl( \p_{\rho_1}
{\rm g}_{\mu_1\nu_1} \p_{\rho_2} {\rm g}_{\nu_2\mu_2} +
\p_{\rho_2} {\rm g}_{\mu_1\nu_1} \p_{\rho_1} {\rm g}_{\nu_2\mu_2} \Bigl)
\end{equation}
and ${\mathbf S}_{\rho_1\rho_2} \equiv ({\rm g}^{-1})^{\mu_2 \mu_1}
{\mathbf S}_{\rho_1\rho_2\mu_1\mu_2}$.
Note that ${\mathbf S}_{\rho_1\rho_2\mu_1\mu_2}$ and ${\mathbf S}_{\rho_1\rho_2}$
are symmetric with respect to $\rho_1 \leftrightarrow \rho_2$.
This map constitutes a generalization of the previous measure
transformation \eq{sw-measure}. Our result is an exact map for the case with
fourth-order derivatives since it includes all powers of gauge fields and the
parameter $\theta$. This identity has been perturbatively checked
up to some nontrivial orders in \ct{derivative-dms,pal,wyllard2}
with perfect agreements.\footnote{\label{boson-derivative}
Here we would like to put forward an interesting
observation. It was well-known \ct{andreev-tseytlin} that the leading derivative
corrections in bosonic string theory start with two-derivatives,
whose exact result including all orders in $F$ was recently obtained
in \ct{andreev}. Thus, if we were adopted the bosonic result with an
assumption that the NC part \eq{nc-dbi-k} were common
for bosonic string and superstring theories (that would be wrong),
we would definitely be on a wrong way.
So the perfect agreement in the identity \eq{sw-derivative-0} is
quite surprising since Eq.\eq{nc-dbi-k} already singles out
the superstring result prior to the bosonic one,
though that was {\it a priori} not clear.
It was also shown in \ct{wyllard2}
that the result in \ct{andreev} for the bosonic string is
not invariant under the SW map. All these seem to imply that the bosonic string
needs to incorporate an effect of tachyons from the outset both in commutative
and in NC descriptions, as suggested in \ct{wyllard2}.
See Mukhi and Suryanarayana in \ct{exact-sw-map} for a relevant discussion.}

Let us consider the next order SW map. Before going to $\CO(\kappa)$
corrections, we have to check to what extent the approximation
\eq{sw-sw-equivalence} could be valid, in other words, what order the leading
derivative corrections, $\Delta \what{S}_{DBI}$,
start with. Since the commutative and NC descriptions arise from the same open
string theory depending on different regularizations, it is natural to expect
that both descriptions share the same structure, namely, the form invariance
\ct{cornalba2}. It is then obvious that the leading correction,
$\Delta \what{S}_{DBI}$, in the NC gauge theory starts with $\CO(\kappa^4)$ as
the commutative one. So we can safely believe the approximation
\eq{sw-sw-equivalence} up to $\CO(\kappa^3)$. Beyond that, we have to take
into account $\Delta \what{S}_{DBI}$ \ct{dbi-derivative}. In order to find
higher order SW map, it is thus enough to expand the metric $\CG^{-1}$
in powers of $\kappa$:
\be \la{expand-sw-metric}
 \CG^{\mu\nu} = \left(\frac{1}{\kappa} \theta {\rm g}^{-1} +
\theta \frac{1}{{\rm g} G {\rm g}^t \theta}
+ \cdots \right)^{\mu\nu}
\ee
with ${\rm g}^t = (1+ \theta F)$. We keep $\theta$ in the second term without
cancelation with the denominator since it will be combined with $F_{\mu\nu}$
in the Riemann tensor to make ${\rm g}_{\mu\nu}$.
After straightforward calculation, we get
\bea \la{sw-derivative-1}
&& \int d^{p+1} y L_\star \Bigl[ \Tr G^{-1}\what{F}(y) W(y, C_k) \Bigr]
\star e^{i k \cdot y} \xx
&=& \int d^{p+1} x \sqrt{\det {\rm g}} \left[ \Tr G^{-1}({\rm g}^{-1}F)
\Bigl(1 - \frac{1}{96}
{\rm g}^{\mu_4 \mu_1}{\rm g}^{\mu_2 \mu_3}(\theta {\rm g}^{-1})^{\rho_4 \rho_1}
(\theta {\rm g}^{-1})^{\rho_2 \rho_3} \mathbf{S}_{\rho_1\rho_2\mu_1\mu_2}
\mathbf{S}_{\rho_3\rho_4\mu_3\mu_4} \right. \xx
&& \left. \hspace{3cm} + \frac{1}{192} (\theta {\rm g}^{-1})^{\rho_4 \rho_1}
(\theta {\rm g}^{-1})^{\rho_2 \rho_3} \mathbf{S}_{\rho_1\rho_2}
\mathbf{S}_{\rho_3\rho_4}  \Bigr) \right. \xx
&& -\frac{1}{24}  {\rm g}^{\mu_2 \mu_3}(\theta {\rm g}^{-1})^{\rho_2 \rho_3}
\left\{ \Bigl( \frac{1}{{\rm g} G {\rm g}^t \theta}
\Bigr)^{\mu_4 \mu_1}(\theta {\rm g}^{-1})^{\rho_4 \rho_1}
 + {\rm g}^{\mu_4 \mu_1} \Bigl(\frac{1}{{\rm g} G {\rm g}^t}
\Bigr)^{\rho_4 \rho_1} \right\}
\mathbf{S}_{\rho_1\rho_2\mu_1\mu_2}
\mathbf{S}_{\rho_3\rho_4\mu_3\mu_4} \xx
&& +\frac{1}{48} (\theta {\rm g}^{-1})^{\rho_2 \rho_3}
\left\{ \Bigl( \frac{1}{{\rm g} G {\rm g}^t \theta}
\Bigr)^{\mu_2 \mu_1}(\theta {\rm g}^{-1})^{\rho_4 \rho_1}
 + {\rm g}^{\mu_2 \mu_1} \Bigl(\frac{1}{{\rm g} G {\rm g}^t}
\Bigr)^{\rho_4 \rho_1} \right\}
\mathbf{S}_{\rho_1\rho_2\mu_1\mu_2}
\mathbf{S}_{\rho_3\rho_4} \\
&& \left. + \frac{1}{24} {\rm g}^{\mu_4 \mu_1} {\rm g}^{\mu_2 \mu_3}
(\theta {\rm g}^{-1})^{\rho_4 \rho_1}
(\theta {\rm g}^{-1})^{\rho_2 \rho_3}
\Bigl( \frac{1}{{\rm g} G {\rm g}^t \theta} \Bigr)^{\nu_1 \nu_2}
\Bigl(\p_{\rho_1} {\rm g}_{\mu_1\nu_1} \p_{\rho_2}
{\rm g}_{\nu_2\mu_2} - \p_{\rho_2}
{\rm g}_{\mu_1\nu_1} \p_{\rho_1} {\rm g}_{\nu_2\mu_2}
\Bigr)\mathbf{S}_{\rho_3\rho_4\mu_3\mu_4} \right. \xx
&& \left. - \frac{1}{48} {\rm g}^{\mu_2 \mu_1}
(\theta {\rm g}^{-1})^{\rho_4 \rho_1}
(\theta {\rm g}^{-1})^{\rho_2 \rho_3}
\Bigl( \frac{1}{{\rm g} G {\rm g}^t \theta} \Bigr)^{\nu_1 \nu_2}
\Bigl(\p_{\rho_1} {\rm g}_{\mu_1\nu_1} \p_{\rho_2}
{\rm g}_{\nu_2\mu_2} - \p_{\rho_2}
{\rm g}_{\mu_1\nu_1} \p_{\rho_1} {\rm g}_{\nu_2\mu_2}
\Bigr)\mathbf{S}_{\rho_3\rho_4} \right]. \nonumber
\eea
We see that the left hand side (and also the first term
on the right hand side) of Eq.\eq{sw-derivative-1} identically
vanishes since $\Tr G^{-1} \what{F} =0$.
Thus the identity \eq{sw-derivative-1} implies that the right hand side must
be a total derivative. We will not check it but leave it as our prediction.
We note that the commutative description in terms of open string variables
can be solely expressed in terms of ${\rm g}_{\mu\nu}$ (after rewriting
${\rm g}^{-1}F= (1-{\rm g}^{-1}) \theta^{-1}$).

More important consequence of Eq.\eq{sw-derivative-1} is the
following. Let us take the metric $G^{\mu\nu}$ out from the
integration on both sides. Since Eq.\eq{sw-derivative-1} is an
identity valid for any arbitrary $G^{\mu\nu}$,
the coefficients of $G^{\mu\nu}$ must be equal too.
Then the left hand side has the form
\be \la{esw-f}
\int d^{p+1} y L_\star \Bigl[\what{F}_{\mu\nu}(y) W(y, C_k) \Bigr]
\star e^{i k \cdot y}.
\ee
We can thus derive the exact SW map of Eq.\eq{esw-f} from
the coefficient of $G^{\mu\nu}$ on the right hand side of Eq.\eq{sw-derivative-1},
up to fourth-order derivative. We will not give the explicit form
since it is rather lengthy but directly readable from Eq.\eq{sw-derivative-1}.
This map has to correspond to the inverse of the exact (inverse) SW
map
\be \la{esw-liu}
F_{\mu\nu}(k) =  \int d^{p+1} y L_\star \Bigl[ \sqrt{\det(1-\theta\what{F})}
\left(\frac{1}{1-\what{F}\theta} \what{F} \right)(y) W(y, C_k) \Bigr]
\star e^{i k \cdot y}
\ee
which was conjectured in \ct{liu} and immediately proved in \ct{exact-sw-map}.

As we argued above, we can continue this procedure using the
expansion \eq{expand-sw-metric} up to $\CO(\kappa^3)$ without
including $\Delta \what{S}_{DBI}$. At each step, we get exact SW
maps including all powers of gauge fields and $\theta_{\mu\nu}$.
Up to our best knowledge, this was never achieved even for
the $\CO(1)$ result \eq{sw-derivative-0}. (But see \ct{{jusch-wess}}
for a formal solution based on the Kontsevich's formality map.)
Such a great simplification is due to the use of the same string
variables using the formula \eq{sw-metric}, originally suggested in \ct{hsy}.
So let us ponder upon possible sources to ruin the
conversion relations \eq{op-cl-gen} and \eq{Gs-gs-gen}.
If quantum corrections are included, the effect of renormalization
group flow of coupling constants might be incorporated into Eq.\eq{Gs-gs-gen}.
But this is only true for asymmetric running of a dilaton field in
commutative and NC theories, which seems not to be the case. Another
source may be a possibility that gauge field dynamics modifies
either $g_{\mu\nu}, G_{\mu\nu}$ or $\theta_{\mu\nu}$, themselves.
As was explained in the previous section and will be shown later,
the dynamics of gauge fields induces the deformation of background
geometry, but this kind of modification is entirely encoded
in ${\rm g}_{\mu\nu}$ or $h_{\mu\nu}$, as indicated in
Eq.\eq{metric-h-g}. Then the variables in Eqs.\eq{op-cl-gen} and
\eq{Gs-gs-gen} in general appear as non-dynamical parameters.
Thus the change of variable \eq{sw-metric} seems to be quite general
independently of gauge field dynamics. If this is so,
we may go much further using the conjectured higher-order
derivative corrections in \ct{wyllard2}.

\section{Deformation Quantization and Emergent Geometry}

In classical mechanics, the set of possible states of a system forms a Poisson
manifold.\footnote{\la{poisson} A Poisson manifold is a differentiable manifold $M$ with
skew-symmetric, contravariant 2-tensor (not necessarily nondegenerate)
$\theta = \theta^{\mu\nu} \p_\mu \wedge \p_\nu \in \Lambda^2 TM$
such that $\{f,g\} = \langle \theta, df \otimes dg \rangle
= \theta^{\mu\nu} \p_\mu f \p_\nu g$
is a Poisson bracket, i.e., the bracket $\{ \cdot , \cdot \}:C^\infty(M)
\times C^\infty(M) \to C^\infty(M)$ is a skew-symmetric bilinear map
satisfying 1) Jacobi identity: $\{f,\{g,h\}\} + \{g,\{h,f\}\}
+ \{h,\{f,g\}\} = 0$ and 2) Leibniz rule:
$\{f,gh\} = g \{f,h\} + \{f,g\}h, \; \forall f,g,h \in C^\infty(M)$.
Poisson manifolds appear as a natural generalization of symplectic
manifolds where the Poisson structure reduces to a symplectic structure if
$\theta$ is nongenerate.}
The observables that we want to measure are the smooth functions
in $C^\infty(M)$, forming a commutative (Poisson) algebra.
In quantum mechanics, the set of possible states is a projective
Hilbert space $\CH$. The observables are self-adjoint operators,
forming a NC C*-algebra. The change from a Poisson
manifold to a Hilbert space is a pretty big one.

A natural question is whether the quantization such as Eq.\eq{poisson-bracket}
for general Poisson manifolds is always possible with a radical change in the
nature of the observables. The problem is how to construct the Hilbert space
for a general Poisson manifold, which is in general highly nontrivial.
Deformation quantization was proposed in \ct{d-quantize} as an alternative,
where the quantization is understood as a deformation of the algebra of
classical observables. Instead of building a Hilbert space from a Poisson
manifold and associating an algebra of operators to it,
we are only concerned with the algebra $\CA$ to deform the commutative
product in $C^\infty(M)$ to a NC, associative product.
In flat phase space such as the case we have considered up to now,
it is easy to show that the two approaches have one to one
correspondence \eq{star-product} through the Weyl-Moyal map \ct{nc-review}.

Recently M. Kontsevich answered the above question in the context of
deformation quantization \ct{kontsevich}. He proved that every
finite-dimensional Poisson manifold $M$ admits a canonical deformation
quantization and that changing coordinates leads to gauge equivalent star
products. We briefly recapitulate his results which will be crucially
used in our discussion.

Let $\CA$ be the algebra over $\IR$ of smooth functions on a
finite-dimensional $C^\infty$-manifold $M$.
A star product on $\CA$ is an associative
$\IR[[\hbar]]$--bilinear product on the algebra $\CA[[\hbar]]$, a formal
power series in $\hbar$ with coefficients in $C^\infty (M) = \CA$,
given by the following formula for $f,g \in \CA \subset \CA[[\hbar]]$:
\be \la{d-star}
(f,g) \mapsto f \star g = fg + \hbar{B_1(f,g)}+ \hbar^2{B_2(f,g)}+ \cdots
\in A[[\hbar]]
\ee
where $B_i(f,g)$ are bidifferential operators.
There is a natural gauge group which acts on star products. This group
consists of automorphisms of $\CA[[\hbar]]$ considered as
an $\IR[[\hbar]]$--module (i.e. linear transformations $\CA \to \CA$
parameterized by $\hbar$). If $D(\hbar) = 1+ \sum_{n\geq{1}}{\hbar^{n}D_n}$
is such an automorphism where $D_n: \CA \to \CA$ are differential operators,
it acts on the set of star products as
\be \la{star-equiv}
\star \rightarrow \star^\prime, \quad
f(\hbar) \star^\prime g(\hbar) = D(\hbar)
\Bigl( D(\hbar)^{-1} (f(\hbar)) \star  D(\hbar)^{-1} (g(\hbar)) \Bigr)
\ee
for $f(\hbar), g(\hbar) \in \CA[[\hbar]]$.
This is evident from the commutativity of the diagram
$$
\xymatrix@+0.5cm{ \CA[[\hbar]] \times \CA[[\hbar]]
\ar[r]^-\star \ar[d]_{D(\hbar) \times D(\hbar)}\ & \CA[[\hbar]]
\ar[d]^{D(\hbar)} \cr
\CA[[\hbar]] \times \CA[[\hbar]] \ar[r]^-{\star^\prime} & \CA[[\hbar]] \cr } \
$$

We are interested in star products up to gauge equivalence.
This equivalence relation is closely related to the cohomological Hochschild
complex of algebra $\CA$ \ct{kontsevich}, i.e.
the algebra of smooth polyvector fields on $M$.
For example, it follows from the
associativity of the product \eq{d-star} that the symmetric part of $B_1$
can be killed by a gauge transformation which is a coboundary in the
Hochschild complex, and that the antisymmetric part of $B_1$,
denoted as $B^-_1$, comes from a bivector field $\a \in \Gamma(M,\Lambda^2
TM)$ on $M$:
\be \la{bi-vector}
B_1^-(f,g) = \langle \a, df \otimes dg \rangle.
\ee
In fact, any Hochschild coboundary can be removed by a gauge transformation
$D(\hbar)$, so leading to the gauge equivalent star product \eq{star-equiv}.
The associativity at $\CO(\hbar^2)$ further constrains
that $\a$ must be a Poisson structure on $M$, in other words,
$[\a,\a]_{SN} = 0$, where the bracket is the Schouten-Nijenhuis bracket
on polyvector fields (see \ct{kontsevich} for the definition of this bracket
and the Hochschild cohomology). Thus, gauge equivalence classes of
star products modulo $\CO(\hbar^2)$ are classified by Poisson structures on $M$.
It was shown \ct{kontsevich} that there are no other obstructions to deforming
the algebra $\CA$ up to arbitrary higher orders in $\hbar$.

For an equivalence class of star products for any Poisson manifold,
Kontsevich arrived at the following general results.

{\it Theorem 1.1} in \ct{kontsevich}: The set of gauge equivalence classes
of star products on a smooth manifold $M$ can be naturally identified
with the set of equivalence classes of Poisson structures depending formally
on $\hbar$
\be \la{theorem-1.1}
\a = \a(\hbar)= \a_1 \hbar + \a_2 \hbar^2 + \cdots
\in \Gamma(M, \Lambda^2 TM)[[\hbar]], \quad
[\a,\a]_{SN} = 0 \in \Gamma(M, \Lambda^3 TM)[[\hbar]]
\ee
modulo the action of the group of formal paths in the diffeomorphism group
of $M$, starting at the identity diffeomorphism.

{\it Theorem 2.3} in \ct{kontsevich}: Let $\a$ be a Poisson bi-vector field in a
domain of $\IR^d$. The formula
\be \la{k-star}
f \star g = \sum_{n=0}^\infty \hbar^n \sum_{\Gamma \in G_n}
w_{\Gamma} B_{\Gamma,\a}(f,g)
\ee
defines an associative product. If we change coordinates, we obtain a gauge
equivalent star product.

The formula $\eq{k-star}$ has a natural interpretation in terms of Feynman
diagrams for the path integral of a topological sigma model \ct{catt-feld}.

The simplest example of a deformation quantization is the Moyal product
\eq{star-product} for the Poisson structure on $\IR^d$ with constant
coefficients $\a^{\mu\nu}= i \theta^{\mu\nu}/2$. If $\a^{\mu\nu}$ are not
constant, a global formula is not yet available but can be perturbatively
computed by the prescription given in \ct{kontsevich}.
Up to the second order, this formula can be written as follows
\bea \la{k-star-2}
f \star g &=& fg+ \hbar{\alpha^{ab}\partial_a{f}\partial_b{g}}
+ \frac{\hbar^2}{2} \alpha^{a_1 b_1}\alpha^{a_2 b_2}
{\partial_{a_1}\partial_{a_2}{f}\partial_{b_1}
\partial_{b_2}{g}} \xx
&& +\frac{\hbar^{2}}{3}{\alpha^{a_1 b_1}\partial_{b_1}\alpha^{a_2b_2}}
(\partial_{a_1}\partial_{a_2}{f}\partial_{b_2}{g}
+\partial_{a_1}\partial_{a_2}{g}\partial_{b_2}{f})
+O(\hbar^{3}).
\eea

Now we are ready to promote the properties such as the Darboux theorem
discussed in section 1 to the framework of deformation quantization.
Since $\w$ and $\w^\prime$ in Eq.\eq{darboux} are related by diffeomorphisms,
according to the {\it Theorem 1.1}, the two star products $\star_\w$
and $\star_{\w^\prime}$ defined by the Poisson structures $\w^{-1}$
and ${\w^\prime}^{-1}$, respectively, should be gauge equivalent.
Conversely, if we make an arbitrary change of coordinates,
$y^\mu \mapsto x^a(y)$, in the Moyal $\star$-product \eq{star-product},
which is nothing but Kontsevich's star product \eq{k-star} with the
constant Poisson bi-vector, we get a new star product defined by a
Poisson bi-vector $\alpha(\hbar)$.
But the resulting star product has to be gauge equivalent
to the Moyal product \eq{star-product} and $\alpha(\hbar)$ should be
determined by the original Poisson bi-vector $\theta^{\mu\nu}$.
This is the general statement of the {\it Theorem 2.3}, which was explicitly
checked by Zotov in \ct{zotov} where he obtained the deformation quantization
formula up to the third order.

We copy the result in \ct{zotov} for completeness and for our later use.
\bea \la{zotov-star}
f \star_M g &=& fg+\hbar{\alpha^{ab}\partial_a {f} \partial_b {g}} \xx
&& +\hbar^2\left[\frac{1}{2} \alpha^{a_1 b_1}\alpha^{a_2 b_2}
{\partial_{a_1}\partial_{a_2}{f} \partial_{b_1}\partial_{b_2}{g}}
 +\frac{1}{3}{\alpha^{a_1 b_1}\partial_{b_1}\alpha^{a_2 b_2}}
(\partial_{a_1}\partial_{a_2}{f}\partial_{b_2}{g}
+\partial_{a_1}\partial_{a_2}{g}\partial_{b_2}{f}) \right] \xx
&& +\hbar^{3}\left[\frac{1}{6}
\alpha^{a_1 b_1}\alpha^{a_2 b_2}\alpha^{{a_3 b_3}}
\partial_{a_1}\partial_{a_2}\partial_{a_3}{f}
\partial_{b_1}\partial_{b_2}\partial_{b_3}{g} \right. \\
&& \qquad + \frac{1}{3}{\alpha^{a_1 b_1} \partial_{b_1} \alpha^{a_2 b_2}
\partial_{b_2} \partial_{a_1} \alpha^{a_3 b_3}}
(\partial_{a_2}\partial_{b_3}{f}\partial_{a_3}{g}
- \partial_{a_2}\partial_{b_3}{g}\partial_{a_3}{f}) \xx
&& \qquad + \Bigl(\frac{2}{3}\alpha^{a_1 b_1}\partial_{b_1}\alpha^{a_2 b_2}
\partial_{b_2}\alpha^{a_3 b_3} +
\frac{1}{3}\alpha^{a_2 b_2}\partial_{b_2}\alpha^{a_1 b_1}
\partial_{b_1}\alpha^{b_3 a_3} \Bigr)
\partial_{a_2}\partial_{b_3}{f}\partial_{a_3}\partial_{a_1}{g} \xx
&& \qquad + \frac{1}{6}{\alpha^{a_1 b_1} \alpha^{a_2 b_2} \partial_{b_1}
\partial_{b_2} \alpha^{a_3 b_3}}
(\partial_{a_1}\partial_{a_2}\partial_{a_3}{f}\partial_{b_3}{g}
- \partial_{a_1} \partial_{a_2} \partial_{a_3}{g}\partial_{b_3}{f}) \xx
&& \qquad \left. +\frac{1}{3}\alpha^{a_1 b_1}\partial_{b_1}\alpha^{a_2 b_2}
\alpha^{a_3 b_3}
(\partial_{a_1}\partial_{a_2}\partial_{a_3} {f}
\partial_{b_2}\partial_{b_3}{g}-
\partial_{a_1}\partial_{a_2}\partial_{a_3}{g}
\partial_{b_2}\partial_{b_3}{f})\right] + \CO(\hbar^{4}) \nonumber
\eea
where\footnote{We scale $\vartheta^{ij} \to \frac{i}{2}
\theta^{\mu\nu}$ in \ct{zotov}
to be compatible with the definition \eq{star-product}
and we denote $\p_a \equiv \frac{\p}{\p x^a}$.}
\bea \la{alpha}
\alpha^{ab} &=& \frac{i}{2} \theta^{\mu\nu}
\frac{\partial x^a}{\partial y^\mu} \frac{\partial{x^b}}{\partial{y^\nu}}+
\frac{\hbar^{2}}{16}\left[-\frac{i}{3}\theta^{\mu_1\nu_1} \theta^{\mu_2\nu_2}
\theta^{\mu_3\nu_3} \frac{\partial^3 x^a}{\partial y^{\mu_1}
\partial y^{\mu_2} \partial y^{\mu_3}}
\frac{\partial^3 x^b}{\partial y^{\nu_1} \partial y^{\nu_2}
\partial y^{\nu_3}} \right. \xx
&& \left.+\frac{2}{9}S^{a_1 a_2
    a_3}\partial_{a_1}\partial_{a_2}\partial_{a_3} \alpha^{ab}
+ \theta^{\mu_1\nu_1}\theta^{\mu_2\nu_2}
\frac{\partial^2 x^{a_1}}{\partial y^{\mu_1} \partial y^{\mu_2}}
\frac{\partial^2 x^{b_1}}{\partial y^{\nu_1} \partial y^{\nu_2}}
\partial_{a_1} \partial_{b_1} \alpha^{ab} \right]+ \CO(\hbar^3)
\eea
and $S^{abc}$ is given by
\be \la{zotov-s}
S^{abc} = \theta^{\mu_1\nu_1} \theta^{\mu_2\nu_2} \left(
\frac{\partial^2 x^a}{\partial y^{\mu_1} \partial y^{\mu_2}}
\frac{\partial x^b}{\partial y^{\nu_1}} \frac{\partial x^c}{\partial
y^{\nu_2}}
+ \frac{\partial^2 x^c}{\partial y^{\mu_1} \partial y^{\mu_2}}
\frac{\partial x^a}{\partial y^{\nu_1}} \frac{\partial x^b}{\partial y^{\nu_2}}
+ \frac{\partial^2 x^b}{\partial y^{\mu_1} \partial y^{\mu_2}}
\frac{\partial x^c}{\partial y^{\nu_1}} \frac{\partial x^a}{\partial
y^{\nu_2}} \right).
\ee
The differential operator in the automorphism \eq{star-equiv}
necessary for obtaining Eqs.\eq{zotov-star} and \eq{alpha} is the following
\be \la{diff-d}
D(\hbar) = 1 +\frac{\hbar^2}{16} \left[\theta^{\mu_1\nu_1}\theta^{\mu_2\nu_2}
\frac{\partial^2 x^a}{\partial y^{\mu_1} \partial y^{\mu_2}}
\frac{\partial^2 x^b}{\partial y^{\nu_1} \partial y^{\nu_2}}
\partial_a \partial_b +\frac{2}{9} S^{abc} \partial_a\partial_b\partial_c \right]
+ \CO(\hbar^3).
\ee
Note that $f \star_M g \equiv D(\hbar) \Bigl( D(\hbar)^{-1} (f)
\star  D(\hbar)^{-1} (g) \Bigr)$ in Eq.\eq{zotov-star} is
the Moyal star product \eq{star-product} but after a change of coordinates it
becomes equivalent to the general Kontsevich star product \eq{k-star-2} up to
the gauge equivalence map \eq{diff-d}, thus checking the {\it Theorem 2.3}.
Also notice that
\bea \la{star-bracket}
[f, g]_\star &\equiv& f \star g - g \star f \xx
&=& 2 \hbar{\alpha^{ab}\partial_a {f} \partial_b {g}} + \CO(\hbar^3)
\eea
since $\CO(\hbar^2)$ is symmetric with respect to $f \leftrightarrow g$.

Since the map \eq{diff-d} is explicitly known, we can now solve
the gauge equivalence \eq{star-equiv}. First let us represent
the coordinates $x^\mu(y)$ as in Eq.\eq{cov-coord} to study
its consequence from gauge theory point of view.
The equivalence \eq{zotov-star} immediately
leads to \ct{jurco-schupp,jusch-wess,wess-js} \footnote{For a comparison with these
literatures, $D(\hbar)^{-1}:\CA_x[[\hbar]] \to \CA_y[[\hbar]]$ is understood
as $\CD \equiv D(\hbar) \circ \rho^*$ and $x^\mu(y) = \CD y^\mu$
in their notation since Eq.\eq{diff-d} is already including
the coordinate transformation $\rho^*$.}
\be \la{esw-deformation}
[x^\mu,x^\nu]_{\star} = i(\theta - \theta \what{F}(y) \theta)^{\mu\nu}
= 2 D(\hbar)^{-1}(\alpha^{\mu\nu})
\ee
where the left hand side is the Moyal product \eq{star-product}.
As a check, one can easily see that Eq.\eq{esw-deformation} is trivially
satisfied if Eqs.\eq{alpha} and \eq{diff-d} are substituted
for the right hand side with $\hbar = 1$.
Note that Eq.\eq{esw-deformation} is an exact result since the higher order
terms in Eq.\eq{star-bracket} identically vanish.

By our construction, the new Poisson structure
\be \la{esw-inverse}
\a^{\mu\nu}(x) = \frac{i}{2}\left(\frac{1}{B+F}\right)^{\mu\nu}(x) =
 \frac{i}{2}(\theta {\rm g}^{-1})^{\mu\nu}(x)
\ee
belongs to the same equivalence class as constant $\theta^{\mu\nu}=(1/B)^{\mu\nu}$,
but now depends on dynamical gauge fields.
Thus, if it is determined how the map $D(\hbar)$ depends on the coordinate
transformations as in Eq.\eq{diff-d}, one can in principle
calculate exact SW maps from Eq.\eq{esw-deformation} up to a desired order.
As it should be, Eq.\eq{esw-deformation} reduces to Eq.\eq{semi-esw}
at the leading order where $D(\hbar) \approx 1$. In general, it
definitely contains derivative corrections coming from the higher-order terms in
$D(\hbar)$.\footnote{The leading derivative corrections calculated from
Eq.\eq{esw-deformation} are four derivative terms consistently with
Eqs.\eq{sw-derivative-1} and \eq{esw-f} which are based on superstring
theory. As was mentioned in the footnote \ref{boson-derivative}, the bosonic
string case starts with two derivative terms. It is not so clear how to
reproduce the bosonic string result \ct{andreev} within the deformation
quantization scheme by incorporating tachyons. It would be an interesting
future work.} Thus the identity \eq{esw-deformation} defines the exact SW map
with derivative corrections and corresponds to a quantum deformation
of Eq.\eq{darboux} or equivalently Eq.\eq{semi-esw}.
Incidentally, we can also get the inverse SW map from Eq.\eq{esw-inverse}
by solving Eq.\eq{alpha} (at least perturbatively) which
is of the form $\a^{\mu\nu}=
\half [x^\mu, x^\nu]_\star +$ terms with derivatives of $\a^{\mu\nu}$.
Thus, getting a full quantum deformation reduces to the calculation of
$\a(\hbar)$ or $D(\hbar)$, as done up to $\CO(\hbar^2)$ in \eq{alpha} and
\eq{diff-d}.

The above construction definitely shows that the
deformation quantization is a NC deformation of the
diffeomorphism symmetry \eq{darboux}.
Since NC gravity is based on a NC deformation of
the diffeomorphism group \ct{nc-gravity,szabo}, we expect the
emergent gravity may be a NC gravity in general. We will find
further evidences for this connection.

As was shown in \ct{liu,jusch-wess}, using the exact
SW map \eq{esw-deformation} together with \eq{op-cl-gen} and \eq{Gs-gs-gen},
it is possible to prove the SW equivalence \eq{equiv-dbi-gen}, or more generally,
Eq.\eq{general-sw-equivalence}. Conversely, we showed in section 2 that
the SW map \eq{esw-deformation} at the leading order directly results from the
SW equivalence \eq{sw-equiv}. As we checked above, Eq.\eq{esw-deformation} is a direct
consequence of the gauge equivalence \eq{star-equiv} between
the star products $\star_{\omega^\prime}$ and $\star_\omega$ defined
by the symplectic forms  $\omega^\prime = B + F(x)$ and $\omega=B$, respectively.
One might thus claim that the SW equivalence \eq{general-sw-equivalence} is
just the statement of the gauge equivalence \eq{star-equiv} between
star products.

We would like to point out some beautiful picture working in these
arguments. First note that symplectic (or more generally Poisson)
structures in a gauge equivalence class are related to each other by
the diffeomorphism symmetry, which is realized as the gauge
equivalence \eq{star-equiv} after deformation as illustrated
in Eq.\eq{zotov-star}. This is precisely the statement of {\it Theorem 1.1}.
We realize from the argument in section 1 that the gauge equivalence
\eq{star-equiv} is also related to the $\Lambda$-symmetry \eq{lambda-symmetry}
where the local deformation of symplectic structure is due to the
dynamics of gauge fields who
live in NC spacetime \eq{nc-spacetime}, as shown in \eq{esw-inverse}.
Thus the dynamics of gauge fields appears as the local deformation
of symplectic structures which always belong to the same gauge
equivalence class, so it can entirely be translated into the
diffeomorphism symmetry according to the {\it Theorem 2.3}.

But notice that not all diffeomorphism does deform the symplectic
structure. For example, if the diffeomorphism is generated by a
vector field $X_\l$ satisfying $\CL_{X_\l} B =0$, i.e. $X_\l \in Ham(M)$,
it does not change the symplectic structure $\w = B$.
Let us recall the argument about the Moser lemma in section 1.
For $\w^\prime = \w + dA$, there is a flow $\phi$ generated by a vector field $X$
such that $\phi^*(\w^\prime) =\w$.
But the gauge transformation $A \to A + d \lambda$ only affects the vector field
as $X \to X + X_\l$ where $\iota_{X_\l} \w + d \l = 0$.
The action of $X_\l$ on a smooth function $f$ is given by $ X_\l(f) = \{ \l,
f\}$ and, upon quantization \eq{poisson-bracket}, $ X_{\what\l}(\what{f}) =
- i [ \what\l, \what{f} ]_\star$,
which is exactly the NC $U(1)$ gauge transformation.

Also note that the gauge equivalence \eq{star-equiv} is defined up to the
following inner automorphism \ct{jurco-schupp}
\be \la{inner-autom}
f(\hbar) \to \l(\hbar) \star f(\hbar) \star \l(\hbar)^{-1}
\ee
or its infinitesimal version is
\be \la{inner-auto-0}
\d f(\hbar) = i [ \l, f ]_\star.
\ee
The above similarity transformation \la{inner-auto} definitely does not
change star products. For $f(\hbar) = x^\mu(y)$, Eq.\eq{inner-auto-0} is equal to
the NC gauge transformation \eq{nc-gt}
with the definition $\l(\hbar) \equiv \what\l$ since
$[y^\mu, \what\l]_\star = i \theta^{\mu\nu} \p_\nu \what\l$.
This is a quantum deformation of Eq.\eq{nc-gauge-tr}.

In consequence, the $U(1)$ gauge symmetry is realized as the
symplectomorphism $Ham(M)$ on a symplectic manifold $M$ and,
upon quantization \eq{d-star}, it appears as the inner
automorphism \eq{inner-autom}, which is the NC $U(1)$ gauge
symmetry \ct{cornalba,jurco-schupp,jusch-wess,wess-js}.

If the $\Lambda$-symmetry \eq{lambda-symmetry} happens to be an exact gauge
symmetry, a puzzle arises. If this is the case, two symplectic structures
$\omega^\prime = B + F(x)$ and $\omega=B$ are related by the local gauge
symmetry \eq{lambda-symmetry} and thus the gauge fields should be physically unobservable.
But we know well that the physical configuration space of (NC) gauge
theory is nontrivial. The puzzle can be resolved by noticing that
the NC spacetime \eq{nc-spacetime} is a background induced by a (homogeneous)
condensation of gauge fields.
Consequently, the $\Lambda$-symmetry is spontaneously
broken to the ordinary $U(1)$ gauge symmetry since the background \eq{spacetime-vacuum}
preserves only the latter. The spontaneous
symmetry breaking \eq{spacetime-vacuum} thus allows us to differentiate
gauge fields fluctuating around the background \eq{spacetime-vacuum}
up to the $U(1)$ gauge symmetry.
This is a usual spontaneous symmetry breaking in quantum field
theory, but for the infinite-dimensional diffeomorphism symmetry \ct{g/h},
since, as we discussed before, the $\Lambda$-symmetry is realized as
the diffeomorphism symmetry $Diff(M)$ via the Darboux theorem
while the $U(1)$ gauge symmetry appears as the symplectomorphism $Ham(M)$.
The symmetry breaking \eq{spacetime-vacuum} therefore explains why gravity
is physically observable in spite of the {\it gauge symmetry} \eq{lambda-symmetry}.

Now we are fairly ready to speculate a whole picture about the
emergent gravity from NC spacetime. The $U(1)$ gauge theory defined
by \eq{dbi-general} respects the $\Lambda$-symmetry \eq{lambda-symmetry}
since the underlying sigma model \eq{string-action} clearly respects this
symmetry. The $\Lambda$-symmetry is mapped to $G=Diff(M)$ via the
Darboux theorem and is realized as the gauge equivalence \eq{star-equiv}
after NC deformation. The ordinary $U(1)$ gauge symmetry appears as
the symplectomorphism $H=Ham(M) \subset Diff(M)$, which is realized as
the NC $U(1)$ gauge symmetry.
But the vacuum spacetime \eq{nc-spacetime} preserves only
the symplectomorphism $H$, so the diffeomorphism symmetry $G$ is
spontaneously broken to $H$. Therefore the physical deformation of
symplectic structures takes values in $G/H$ or more precisely the
quantum deformation of $G/H$, which is equivalent to the gauge orbit
space of NC gauge fields or the physical configuration space of NC
gauge theory. (In general, there can be a large variety of $G/H$
with different topology for the gauge equivalence class defined by Eq.\eq{star-equiv},
which might be detected by the Hochschild (or Chevally) cohomology \ct{kontsevich}.)
According to the symmetry breaking $G \to H$,
the dynamical fields $x^\mu(y)$ in \eq{cov-coord}
define a quantum deformation from $y^\mu$,
vacuum expectation values specifying the background \eq{spacetime-vacuum},
along a vector field $X \in LDiff(M)$, Lie algebra of $Diff(M)$.
But the gauge symmetry \eq{inner-autom} introduces an equivalence relation
between the dynamical coordinates $x^\mu(y)$.
Thus the embedding functions $x^\mu(y)$ subject to the
equivalence relation $x^\mu \sim x^\mu + \d x^\mu$ coordinatize
the quotient space $G/H$.

According to the Goldstone's theorem \ct{goldstone} for the symmetry breaking $G \to
H$, massless particles, the so-called Goldstone bosons should appear
which are dynamical variables taking values in the quotient space
$G/H$. Since $G$ is the diffeomorphism symmetry, we assert that the
order parameter emerging from a nonlinear realization $G/H$ should be
in general spin-2 gravitons \ct{g/h}. According to the conjecture,
the gravitational fields $e^\a_\mu (y)$ in Eq.\eq{metric-h-g}
might be identified with the Goldstone bosons for the spontaneous
symmetry breaking \eq{spacetime-vacuum}.
We already gave a supporting argument in section 1
that the dynamics of NC gauge fields appears as the fluctuation of
geometry through general coordinate transformations in $G =
Diff(M)$. We will see that a NC gauge theory describes an emergent geometry
in the way that the fluctuation of gauge fields in NC spacetime \eq{nc-spacetime}
induces a deformation of the vacuum manifold, e.g. $\IR^d$ for constant
$\theta^{\mu\nu}$.

It should be very important to completely determine the structure of
emergent gravity based on the framework of the nonlinear
realization $G/H$ \ct{g/h} (including a full quantum deformation).
Unfortunately this goes beyond the present scope.
Instead we will confirm the conjecture by
considering the self-dual sectors for ordinary and NC gauge theories.
We will see that so beautiful structures about gravity, e.g. the twistor
space \ct{penrose}, naturally emerge from this construction.
Since the emergent gravity seems to be very generic if the conjecture is true anyway,
we believe that the correspondence between self-dual NC electromagnetism and
self-dual Einstein gravity is enough to strongly guarantee the conjecture.

\section{NC Instantons and Gravitational Instantons}

To illustrate the correspondence of NC gauge theory with gravity, we will
explore in this section the equivalence found in \ct{sty,ys}
between NC $U(1)$ instantons and gravitational instantons.
To make the essence of emergent gravity clear as much as possible,
we will neglect the derivative corrections and consider
the usual NC description with $\Phi=0$. The semi-classical
approximation, or slowly varying fields, means that the Moyal star
product \eq{star-product} is approximated only to the first order,
$\CO(\theta)$, in which the NC field strength \eq{nc-f} is replaced by
Eq.\eq{semi-nc-f}. In next section we will consider the effect of
derivative corrections using the background independent formalism of
NC gauge theory \ct{sw,seiberg}, namely, with $\Phi = -B$.
This section will be mostly a mild extension of the previous works \ct{sty,ys} with
more focus on the emergent gravity and the relation to the twistor space.

Let us consider electromagnetism in the NC spacetime \eq{nc-spacetime}.
The action for the NC $U(1)$ gauge theory in flat Euclidean $\IR^4$
is given by
\begin{equation}\label{nced}
\widehat{S}_{\mathrm{NC}} = \frac{1}{4}\int\! d^4 y \,\widehat{F}_{\mu\nu}
\star \widehat{F}^{\mu\nu}.
\end{equation}
Contrary to ordinary electromagnetism, the NC $U(1)$ gauge theory admits
non-singular instanton
solutions satisfying the NC self-duality equation \ct{nek-sch},
\be \la{nc-self-dual}
{\widehat F}_{\mu\nu} (y) = \pm \half
\varepsilon_{\mu\nu\lambda\sigma}{\widehat F}_{\lambda\sigma} (y).
\ee

When we consider NC instantons, the ADHM construction depends only
on the combination $\mu^a = \theta^{\mu\nu}\eta^{(\pm)a}_{\mu\nu}$ \cite{sw,kly}
for anti-self-dual (ASD) (with + sign) and self-dual (SD)
(with - sign) instantons where
$\eta^{(+) a}_{\mu\nu} = \eta^a_{\mu\nu}$ and $\eta^{(-) a}_{\mu\nu} =
{\bar \eta}^a_{\mu\nu}$ are three $ 4 \times 4$ SD and
ASD 't Hooft matrices \ct{sty}. If the instanton is ASD in the NC spacetime
satisfying $\theta^{\mu\nu}\eta^{(-)a}_{\mu\nu}=0$, the
ADHM equation then gets a nonvanishing deformation, which puts a non-zero minimum
size of NC instantons. In this case, the small instanton singularities are
eliminated and the instanton moduli space is thus non-singular \ct{nek-sch}.
However, if the instanton is SD, the deformation is vanishing.
Thus the small instanton singularity is not
eliminated and the instanton moduli space is still singular.
The so-called localized instantons in this case are
generated by shift operators \cite{minwalla}.

As was explained in section 2, the NC gauge theory \eq{nced} has an
equivalent dual description through the SW map in terms of ordinary gauge
theory on commutative spacetime \ct{sw}.
Applying the maps \eq{semi-esw} and \eq{sw-measure} to the action \eq{nced},
one can get the commutative nonlinear electrodynamics \ct{hsy,ban-yang}
equivalent to Eq.\eq{nced} in the semi-classical
approximation,\footnote{Here we would like to correct an incorrect statement,
{\it Proposition 3.1} in \ct{milano}, to remove a disagreement
with existing literatures, especially, with \ct{fidanza}.
See the comments in page 11.
The {\it Proposition 3.1} states that the terms of order $n$ in $\theta$
in the NC Maxwell action \eq{nced} via SW map form a homogeneous polynomial
of degree $n+2$ in $F$ without derivatives of $F$.
The proposition is also inconsistent with our general
result about derivative corrections in section 2 and 3. This
disagreement was recently pointed out in \ct{milano-correction}.

The proposition was based on
a wrong observation that the derivation acting on the $\theta$'s appearing in
star products always gives rise to total derivatives. That is not
true in general. For example, let us consider the
following derivation with respect to $\theta^{\mu\nu}$:
$$ \frac{\delta}{\delta \theta^{\mu\nu}} (f \star g \star h)(y)  =
i \Bigl( \p_{[\mu} f \star \p_{\nu]}g \star h +
\p_{[\mu} f \star g \star \p_{\nu]}h + f \star \p_{[\mu} g \star \p_{\nu]}h
\Bigr)(y),$$ where $f,g,h \in C^\infty(M)$ are rapidly decaying functions at
infinity and are assumed to be $\theta$-independent.
The above derivation cannot be rewritten as a total
derivative. If it were the case, it would definitely imply a wrong result:
$ \int d^4 y  (f \star g \star h)(y) =  \int d^4 y  (f \, g \, h)(y)$. This is
not true for the triple or higher multiple star product.}
\begin{equation}\label{ced-sw}
S_{\mathrm{C}} = \frac{1}{4} \int d^4 x \sqrt{\det{{\rm g}}} \;
{\rm g}^{\mu \lambda} {\rm g}^{\sigma\nu} F_{\mu\nu}
F_{\lambda\sigma},
\end{equation}
where the effective metric ${\rm g}_{\mu\nu}$ \ct{rivelles} was defined in
Eq.\eq{effective-metric}.
It was shown in \ct{sty} that the self-duality equation for the action
$S_{\mathrm{C}}$ is given by
\be \la{c-self-dual}
{\bf F}_{\mu\nu} (x) = \pm \half
\varepsilon_{\mu\nu\lambda\sigma}{\bf F}_{\lambda\sigma} (x),
\ee
with the definition \eq{bold-f}. Note that Eq.\eq{c-self-dual} is
nothing but the exact SW map \eq{semi-esw} of the NC self-duality
equation \eq{nc-self-dual}.

A general strategy was suggested in \ct{sty} to solve the self-duality
equation \eq{c-self-dual}.
To be specific, consider self-dual NC $\IR^4$, i.e.,
$\theta^{\mu\nu}{\bar \eta}^a_{\mu\nu}=0$, with the canonical
form $\theta^{\mu\nu} = \frac{\zeta}{2} \eta^3_{\mu\nu}$.
Take a general ansatz for the SD ${\bf F}^+_{\mu\nu}$
and the ASD ${\bf F}^-_{\mu\nu}$ as follows
\be \la{bff-ansatz}
{\bf F}^{\pm}_{\mu\nu}(x) = f^a(x) \eta^{(\pm) a}_{\mu\nu},
\ee
where $f^a$'s are arbitrary functions.
Then the equation \eq{c-self-dual} is automatically satisfied.
Next, solve the field strength $F_{\mu\nu}$ in terms of ${\bf F}^\pm_{\mu\nu}$:
\begin{equation}\label{f-fatf}
F_{\mu\nu} (x) = \Bigl(\frac{1}{1-{\bf F}^\pm \theta}{\bf F}^\pm \Bigr)_{\mu\nu}(x).
\end{equation}
Substituting the ansatz \eq{bff-ansatz} into Eq.\eq{f-fatf}, we get
\bea \la{field-sd-f}
&& F_{\mu\nu} = \frac{1}{1-\phi}f^a \eta^a_{\mu\nu} +
\frac{2\phi}{\zeta (1-\phi)}\eta^3_{\mu\nu},
\quad \rm{for \; SD \; case}, \\
\la{field-asd-f}
&& F_{\mu\nu} = \frac{1}{1-\phi}f^a \bar{\eta}^a_{\mu\nu} -
\frac{2\phi}{\zeta (1-\phi)}\eta^3_{\mu\nu}, \quad \rm{for \; ASD \; case},
\eea
where $\phi \equiv \frac{\zeta^2}{4}\sum_{a=1}^3 f^a(x)f^a (x).$\footnote{
One can rigorously show that the smooth function $\phi$
for the ASD case (\ref{field-asd-f}) satisfies the inequality, $0 \le \phi < 1$.
The proof is done by noticing that
$$ \half \varepsilon^{\mu\nu\alpha\beta} \sqrt{\det{\mathrm{g}}} ({\mathrm{g}}^{-1}
 F)_{\mu\nu} ({\mathrm{g}}^{-1} F)_{\alpha\beta} = -
\frac{16 \phi}{\zeta^2(1- \phi)} $$
since the left-hand side is negative definite
unless zero and $\phi$ is definitely non-negative.}

For the ASD case \eq{field-asd-f}, we get the instanton equation in \ct{sw}
(see also \ct{nonlinear-instanton})
\be \la{sw-instanton}
F_{\mu\nu}^+ \equiv \half(F_{\mu\nu}+ \half \varepsilon_{\mu\nu\rho\sigma}
F_{\rho\sigma}) = \frac{1}{4}(F\widetilde{F}) \theta_{\mu\nu}^+
\ee
since
\be \la{inst-density}
F\widetilde{F} \equiv \half
\varepsilon^{\mu\nu\rho\sigma}F_{\mu\nu}F_{\rho\sigma}=
- \frac{16 \phi}{\zeta^2 (1-\phi)},
\ee
while, for the SD case \eq{field-sd-f},
\be \la{sd-instanton}
F_{\mu\nu}(x) = \half \varepsilon_{\mu\nu\rho\sigma} F_{\rho\sigma}(x).
\ee
Interestingly, using the inverse metric
$$ ({\rm g}^{-1})^{\mu\nu} = \frac{1}{\sqrt{\det{{\rm g}}}}
\Bigl( \half {\rm g}_{\l\l} \delta_{\mu\nu} - {\rm g}_{\mu\nu}\Bigr),$$
Eq.\eq{sw-instanton} can be rewritten as the self-duality
in a curved space described by the metric ${\rm g}_{\mu\nu}$
\be \la{curved-self-dual}
F_{\mu\nu} (x) = - \half \frac{\varepsilon^{\lambda\sigma\rho\tau}}
{\sqrt{\det{{\rm g}}}} {\rm g}_{\mu\lambda}{\rm g}_{\nu\sigma}
F_{\rho\tau} (x).
\ee
It is interesting to compare this with the SD case \eq{sd-instanton}.
It should be remarked, however, that the self-duality in \eq{curved-self-dual}
cannot be interpreted as a usual self-duality equation
in a fixed background since the four-dimensional metric used to define
Eq.\eq{curved-self-dual} depends in turn on the $U(1)$ gauge fields.

It is well-known that there is no nontrivial solution to (A)SD
equation in ordinary $U(1)$ gauge theory.
Since the SD instanton satisfies Eq.(\ref{sd-instanton}),
the exact SW map of localized instantons is thus either trivial or very singular.
This result is consistent with \cite{hashimoto-ooguri}.
Fron now on, we thus focus on the ASD instantons.

Since the field strength \eq{field-asd-f} is given by a (locally) exact
two-form, i.e., $F=dA$, we impose the Bianchi identity for $F_{\mu\nu}$,
\be \la{bianchi}
\varepsilon_{\mu\nu\rho\sigma}\partial_{\nu} F_{\rho\sigma} = 0.
\ee
In the end Eq.\eq{bianchi} leads to general differential equations
governing $U(1)$ instantons \ct{sty}.
The equation \eq{bianchi} was explicitly solved in \ct{sty,sw}
for the single instanton case.
It was found in \ct{sty} that the effective metric \eq{effective-metric} for the
single $U(1)$ instanton is related to the Eguchi-Hanson (EH) metric
\ct{eguchi-hanson}, the simplest asymptotically locally Euclidean (ALE) space,
given by
\begin{equation} \label{eguchi-hanson}
ds^2 = \Bigl(1-\frac{t^4}{\varrho^4}\Bigr)^{-1} d\varrho^2 +
\varrho^2(\sigma_x^2 + \sigma_y^2) + \varrho^2
\Bigl(1-\frac{t^4}{\varrho^4}\Bigr) \sigma_z^2
\end{equation}
where $\sigma_i$ are the $SU(2)$ left-invariant 1-forms satisfying
$d\sigma_i + \varepsilon_{ijk}\sigma_j \wedge \sigma_k=0$.
The metric (\ref{eguchi-hanson}) can be transformed to the K\"ahler metric form
(4.2) in \ct{sty} by the following
coordinate transformation \cite{gibb-pope}:
\begin{eqnarray} \label{transform-eh}
&& r^2(\sigma_x^2 + \sigma_y^2) =  |dz_1|^2 + |dz_2|^2 -
r^{-2} |{\bar z}_1 dz_1 + {\bar z}_2 dz_2 |^2, \nonumber \\
&& r^2 \sigma_z^2  = - \frac{1}{4r^2} ({\bar z}_1 dz_1 + {\bar z}_2 dz_2 -
z_1 d{\bar z}_1 - z_2 d{\bar z}_2 )^2,
\end{eqnarray}
where
\begin{equation}
\varrho^4 = r^4 + t^4
\end{equation}
and $r^2=|z_1|^2 + |z_2|^2$ is the embedding coordinate in field theory.

The EH metric (\ref{eguchi-hanson}) has a curvature that reaches a
maximum at the `origin' $\varrho = t$, falling away to zero in all four directions as
the radius $\varrho$ increases. The apparent singularity in
Eq.(\ref{eguchi-hanson}) at $\varrho = t$ (which is the same singularity
appearing at $r=0$ in the instanton solution constructed in \ct{sty,sw})
is only a coordinate singularity,
provided that $\psi$ is assigned the period $2\pi$ rather than $4 \pi$
(where $\sigma_z = \frac{1}{2}(d\psi + \cos \theta d \phi))$.
Since the radial coordinate runs down only as far as
$\varrho = t$, there is a minimal 2-sphere ${\bf S}^2$ of radius $t$ described
by the metric $t^2(\sigma_x^2 + \sigma_y^2)$.
This degeneration of the generic three dimensional orbits to the two
dimensional sphere is known as a `bolt' \cite{gibbons-hawking}.
As we mentioned above, the NC parameter $\zeta$ in the gauge theory settles
the size of NC $U(1)$ instantons and removes the singularity of instanton
moduli space coming from small instantons.
The parameter $\zeta$ is related to the parameter $t^2$
in the EH metric (\ref{eguchi-hanson}) as $t^2 = \zeta {\widetilde t}^2$
with a dimensionless constant ${\widetilde t}$ and so
to the size of the `bolt' in the gravitational instantons \cite{sty}.
Unfortunately, since $\varrho = t$ corresponds to the origin
$r=0$ of the embedding coordinates, this nontrivial topology is
not visible in the gauge theory description, as was pointed
out in \cite{braden-nekrasov}. However, we see that the dynamical approach
where a manifold is emerging from dynamical gauge fields,
as in Eq.(\ref{induced-metric}), reveals
the nontrivial topology of the D-brane submanifold.

It would be useful to briefly summarize the work \cite{braden-nekrasov}
since it seems to be very related to ours although explicit solutions
are different from each other (see section 4.2 in \cite{braden-nekrasov}).
Braden and Nekrasov constructed $U(1)$ instantons using the deformed ADHM equation
defined on a {\it commutative} space $X$. They showed that the resulting gauge fields
are singular unless one changes the topology of the spacetime and that the $U(1)$
gauge field can have a non-trivial instanton charge if the spacetime
contains non-contractible two-spheres. They thus argued that $U(1)$
instantons on NC $\IR^4$ correspond to non-singular $U(1)$ gauge fields
on a commutative K\"ahler manifold $X$ which is a blowup of $\IC^2$ at a
finite number of points. Also they speculated that the manifold $X$
for instanton charge $k$ can be viewed as a spacetime foam with $b_2 \sim k$.

Now let us show the equivalence between $U(1)$ instantons
in NC spacetime and gravitational instantons \ct{ys}.
In other words, Eq.\eq{nc-self-dual} or Eq.\eq{c-self-dual} describes
gravitational instantons obeying the SD equations \ct{g-instanton}
\be \la{g-instanton}
R_{abcd} = \pm \half \varepsilon_{abef}
{R^{ef}}_{cd},
\ee
where $R_{abcd}$ is a curvature tensor.
The instanton equation \eq{sw-instanton} can be rewritten using
the metric \eq{effective-metric} as follows
\bea \la{self-dual-metric}
&& {\rm g}_{13} = {\rm g}_{24}, \quad  {\rm g}_{14} = - {\rm g}_{23}, \xx
&& {\rm g}_{\mu\mu} = 4 \sqrt{\det{{\rm g}_{\mu\nu}}}
\eea
with $\sqrt{\det{{\rm g}_{\mu\nu}}} = {\rm g}_{11}{\rm g}_{33}-({\rm g}_{13}^2 +
{\rm g}_{14}^2)$ and ${\rm g}_{12} = {\rm g}_{34} =0$ identically.
We will show that Eq.\eq{self-dual-metric} reduces to the so-called
complex Monge-Amp\`ere equation \ct{yau} or the Pleba\~nski equation \ct{plebanski},
which is the Einstein field equation for a K\"ahler metric \ct{ys}.

To proceed with the K\"ahler geometry, let us introduce the complex
coordinates and the complex gauge fields
\bea \la{complex-r4}
&& z_1 = x^2 + i x^1, \qquad z_2 = x^4 + i x^3, \\
\la{complex-a}
&& A_{z_1} = A^2 - i A^1, \quad A_{z_2} = A^4 - i A^3.
\eea
In terms of these variables, Eq.\eq{sw-instanton} are written as
\bea \la{holomorphic}
&& F_{z_1 z_2}  = 0 = F_{\zbar_1 \zbar_2}, \\
\la{stability}
&& F_{z_1 \zbar_1} + F_{z_2 \zbar_2 } = - \frac{i\zeta}{4} F\widetilde{F},
\eea
where $F\widetilde{F} = -4 ( F_{z_1 \zbar_1} F_{z_2 \zbar_2 }
+  F_{z_1 \zbar_2} F_{\zbar_1 z_2}).$
Note that Eq.\eq{holomorphic} is the condition for a holomorphic vector bundle,
but the so-called stability condition \eq{stability} is deformed by
noncommutativity. (See Chap.15 in \ct{gsw-book}.)

One can easily see that the metric ${\rm g}_{\mu\nu}$ is a Hermitian metric \ct{ys}.
That is,
\be \la{hermitian-metric}
ds^2 = {\rm g}_{\mu\nu} dx^\mu dx^\nu = g_{i\bar{j}} dz_i
d\bar{z}_j, \quad i,j=1,2.
\ee
If we let
\be \la{kahler-form}
\varpi = \frac{i}{2} g_{i\bar{j}} dz_i \wedge d\bar{z}_j
\ee
be the K\"ahler form, then the K\"ahler condition is $d \varpi = 0$,
or, for all $i,j,k$,
\be \la{kahler-condition}
\frac{\partial g_{i\bar{j}}}{\partial z^k} = \frac{\partial g_{k\bar{j}}}{\partial z^i}.
\ee
The K\"ahler condition \eq{kahler-condition} is then equivalent to the Bianchi identity
\eq{bianchi} since
\be \la{omega}
\varpi = - (dx^1 \wedge dx^2 + dx^3 \wedge dx^4) + \frac{\zeta}{2} F.
\ee
Thus the metric $g_{i\bar{j}}$ is a K\"ahler metric and
thus we can introduce a K\"ahler potential $K$ defined by
\be \la{kahler-potential}
g_{i\bar{j}} = \frac{\partial^2 K}{\partial z^i \partial \zbar^j}.
\ee
The K\"ahler potential $K$ is related to the integrability condition
of Eq.\eq{holomorphic} (defining a holomorphic vector bundle):
\be \la{complex-gauge}
A_{z_i} = 0, \quad A_{\zbar_i} = 2i \partial_{\zbar_i} (K - \zbar_k z_k).
\ee

Let us rewrite ${\rm g}_{\mu\nu}$ as
\be \la{new-metric}
{\rm g}_{\mu\nu}= \half(\delta_{\mu\nu} + \widetilde{{\rm g}}_{\mu\nu}).
\ee
Then, from Eq.\eq{self-dual-metric}, one can easily see that
\be \la{determinant=1}
 \sqrt{\det{\widetilde{{\rm g}}_{\mu\nu}}} = 1.
\ee
Note that the metric $\widetilde{{\rm g}}_{\mu\nu}$ is also a K\"ahler metric:
\be \la{new-kahler}
\widetilde{g}_{i\bar{j}} = \frac{\partial^2 \widetilde{K}}{\partial z^i \partial \zbar^j}.
\ee
The relation $\det{\widetilde{{\rm g}}_{\mu\nu}} = (\det{\widetilde{g}_{i\bar{j}}})^2$
definitely leads to the Ricci-flat condition
\be \la{monge-ampere}
\det{\widetilde{g}_{i\bar{j}}} = 1.
\ee
Therefore the metric $\widetilde{{\rm g}}_{\mu\nu}$ is both Ricci-flat and
K\"ahler, which is the case of gravitational instantons \ct{ys}.
For example, if one assumes
that $\widetilde{K}$ in Eq.\eq{new-kahler} is a function solely of $r^2 = |z_1|^2 + |z_2|^2$,
Eq.\eq{monge-ampere} can be integrated to give \ct{gibb-pope}
\be \la{pot-k}
\widetilde{K} = \sqrt{r^4 + t^4} + t^2 \log \frac{r^2}{\sqrt{r^4 + t^4} + t^2}.
\ee
This leads precisely to the EH metric \eq{eguchi-hanson} after the coordinate
transformation \eq{transform-eh}.
We thus confirmed that the instanton equation \eq{sw-instanton} is equivalent
to the Einstein field equation for K\"ahler metrics.

The above arguments can be elegantly summarized as the hyper-K\"ahler
condition in the following way \ct{ys}.
Let us consider the line element defined by the metric $\widetilde{{\rm g}}_{\mu\nu}$
\be \la{dia-metric}
ds^2 = \widetilde{{\rm g}}_{\mu\nu} dx^\mu dx^\nu \equiv
 \widetilde{\sigma}_\mu \otimes  \widetilde{\sigma}_\mu.
\ee
It is easy to check that $\widetilde{\sigma}_1 \wedge  \widetilde{\sigma}_2 \wedge
\widetilde{\sigma}_3 \wedge  \widetilde{\sigma}_4 = d^4 x$, in other words,
$\sqrt{\det{\widetilde{{\rm g}}_{\mu\nu}}} = 1$.
We then introduce the triple of K\"ahler forms as follows,
\be \la{unit-2-form}
\widetilde{\omega}^a = \half \eta^a_{\mu\nu} \widetilde{\sigma}^\mu
\wedge \widetilde{\sigma}^\nu, \qquad a=1,2,3.
\ee
One can easily see that
\bea \la{3-kahlerform}
&& \omega \equiv \widetilde{\omega}^2 + i \widetilde{\omega}^1 = dz_1 \wedge dz_2,
\quad \bar{\omega} \equiv \widetilde{\omega}^2 - i \widetilde{\omega}^1
=  d\zbar_1 \wedge d\zbar_2, \xx
&& \Omega \equiv - \widetilde{\omega}^3 = \frac{i}{2}(dz_1 \wedge d\zbar_1
+ dz_2 \wedge d\zbar_2) + \zeta F.
\eea
It is obvious that $d\widetilde{\omega}^a = 0, \; \forall a$. This means that
the metric $\widetilde{{\rm g}}_{\mu\nu}$ is hyper-K\"ahler \ct{ys}, which is
an equivalent statement as Ricci-flat K\"ahler in four dimensions.

Eq.\eq{3-kahlerform} shows how dynamical gauge fields living in NC
spacetime induce a deformation of background geometry
through gravitational instantons, thus realizing the emergent geometry we
claimed before. We see that, if we turn off either gauge fields
or noncommutativity (to be precise, a commutative limit $\zeta \to 0)$,
we simply arrive at flat $\IR^4$.
But, if we turn on both gauge fields and noncommutativity,
the background geometry, say flat $\IR^4$, is nontrivially deformed
and we arrive at a curved manifold.
For instance, hyper-K\"ahler manifolds emerge from NC instantons.
Actually this picture also implies that the flat $\IR^4$ has to be interpreted
as emergent from the homogeneous gauge field
condensation \eq{spacetime-vacuum} \ct{mine}.

$\IR^4$ is the simplest hyper-K\"ahler manifold, viewed as the quaternions
$\IH \simeq \IC^2$. Hyper-K\"ahler manifold is a manifold equipped with
infinitely many (${\bf S}^2$-family) of K\"ahler structures.
This ${\bf S}^2$-family corresponds to the number of inequivalent choices of
local complex structures on $\IR^4$. Since there is no preferred complex
structure, it is democratic to consider $\IR^4 \simeq \IC^2$
with the set of all possible local complex structures simultaneously
at each point, in other words, a $\IP^1={\bf S}^2$ bundle over $\IR^4$.
The total space of this $\IP^1$ bundle is the twistor
space $\CZ$ \ct{penrose}. Since gravitational instantons are also
hyper-K\"ahler manifolds, they also carry a $\IP^1$-family of K\"ahler structures.
So one can similarly construct the corresponding twistor space $\CZ$ describing
curved self-dual spacetime as a $\IP^1$ bundle over a hyper-K\"ahler
manifold $\CM$ \ct{penrose,atiyah}. The twistor space $\CZ$ may also be viewed
as a fiber bundle over $\IP^1$ with a fiber being $\CM$.

Now we will show that the equivalence of NC instantons with
gravitational instantons perfectly fits with the geometry of the
twistor space describing curved self-dual spacetime.
This construction, which closely follows the
results on $N=2$ strings \ct{ooguri-vafa-mpl,ooguri-vafa}, will clarify
how the deformation of symplectic (or K\"ahler) structure on $\IR^4$
due to the fluctuation of gauge fields appears as that of complex structure
of the twistor space $\CZ$ \ct{mine}.

Consider a deformation of the holomorphic (2,0)-form
$\omega = dz_1 \wedge dz_2$ as follows
\be \la{deformation}
\Psi (t) = \omega + i t \Omega + \frac{t^2}{4} \bar{\omega}
\ee
where the parameter $t$ takes values in $\IP^1$. Note that $\Omega$ is a (1,1)
form because of Eq.\eq{holomorphic}.
One can easily see that $d\Psi(t) = 0$ due to the Bianchi identity $dF=0$
and
\be \la{psi-psi}
\Psi (t) \wedge \Psi (t) = 0
\ee
since Eq.\eq{psi-psi} is equivalent to Eq.\eq{sw-instanton}.
Since the two-form $\Psi(t)$ is closed and degenerate, the Darboux theorem
asserts that one can find a $t$-dependent
map $(z_1,z_2)  \to (Z_1(t;z_i, \zbar_i), Z_2(t;z_i, \zbar_i))$ such that
\be \la{symplectic-form}
\Psi (t) = dZ_1(t;z_i,\zbar_i) \wedge dZ_2(t;z_i, \zbar_i).
\ee

When $t$ is small, one can solve \eq{symplectic-form} by expanding
$Z_i(t;z,\zbar)$ in powers of $t$ as
\be \la{small-expansion}
Z_i(t;z,\zbar) = z_i + \sum_{n=1}^{\infty} \frac{t^n}{n}p_n^i(z,\zbar).
\ee
By substituting this into Eq.\eq{deformation}, one gets at ${\cal O}(t)$
\bea \la{exp-eq1}
&& \partial_{z_i} p_1^i = 0, \\
\la{exp-eq2}
&& \epsilon_{ik} \partial_{\zbar_j} p_1^k dz^i \wedge d\zbar^j= i \Omega.
\eea
Eq.\eq{exp-eq1} can be solved by setting $p_1^i = 1/2 \epsilon^{ij}
\partial_{z_j} {\widetilde K}$ and then $\Omega = i/2  \partial_i
{\bar \partial_j}  {\widetilde K} dz^i \wedge d\zbar^j$.
The real-valued smooth function ${\widetilde K}$ is the K\"ahler
potential of $U(1)$ instantons in Eq.\eq{new-kahler}.
In terms of this K\"ahler two-form $\Omega$, Eq.\eq{psi-psi} leads
to the complex Monge-Amp\`ere or the Pleba\'nski equation,
Eq.\eq{monge-ampere},
\be \la{cma-pleb}
\Omega \wedge \Omega = \half \omega \wedge \bar{\omega},
\ee
that is, $\det(\partial_i {\bar \partial_j} {\widetilde K}) =1$.

When $t$ is large, one can introduce another Darboux coordinates
${\widetilde Z}_i(t;z_i,\zbar_i)$ such that
\be \la{symplectic-form2}
\Psi (t) = t^2 d{\widetilde Z}_1(t;z_i,\zbar_i) \wedge
d{\widetilde Z}_2(t;z_i, \zbar_i)
\ee
with expansion
\be \la{small-expansion2}
{\widetilde Z}_i(t;z,\zbar) = \zbar_i + \sum_{n=1}^{\infty} \frac{t^{-n}}{n}
{\widetilde p}_n^i(z,\zbar).
\ee
One can get the solution \eq{deformation} with ${\widetilde p}_1^i = - 1/2
\epsilon^{ij} \partial_{\zbar_j} K$ and $\Omega =
i/2  \partial_i {\bar \partial_j} K dz^i \wedge d\zbar^j$.

The $t$-dependent Darboux coordinates $Z_i(t;z,\zbar)$ and
${\widetilde Z}_i(t;z,\zbar)$ correspond to holomorphic coordinates
on two local charts, where the 2-form $\Psi(t)$ becomes the holomorphic
(2,0)-form, of the dual projective twistor space $\CZ$
as a fiber bundle over ${\bf S}^2$
with a fiber ${\cal M}$, a hyper-K\"ahler manifold.
Here we regard $t$ as a parameter of deformation of complex structure
on ${\cal M}$. The coordinate charts can be consistently glued together
along the equator on $\IP^1$ as $(Z^\prime, t^\prime)
= (t^{-1} f(t;Z), - t^{-1})$ \ct{penrose} and so the complex structure is
extended over $\CZ$. Therefore the Darboux coordintates are related
by a $t$-dependent symplectic transformation on an overlapping
coordinate chart as $f_i(t;Z(t))=t {\widetilde Z}_i(t)$ \ct{ooguri-vafa}.
In this way, the complex geometry of the twistor
space $\CZ$ encodes all the information about the K\"ahler geometry of
self-dual 4-manifolds $\CM$ emerging from NC gauge fields.

This twistor construction clarifies the nature
of emergent gravity; the gauge fields
act as a deformation of the complex structure of twistor
space or the K\"ahler structure of self-dual 4-manifold. In
this way gauge fields in NC spacetime manifest themselves
as a deformation of background geometry, which
is consistent with the picture observed below Eq.\eq{3-kahlerform}.
Thus we should think of the twistor space as already incorporating
the backreaction of NC instantons. This picture is remarkably
similar to that in  \ct{twistor-string} where placing D1-branes
(as instantons in gauge theory) in twistor space is interpreted
as blowing up points in four dimensions dubbed as spacetime
foams and the K\"ahler blowups in four dimensions
are encoded in the twistor space as the backreaction of
the D1-branes. Under the twistor correspondence, each $\IP^1$ in $\CZ$
corresponds to a point on $\CM$.
In particular, D1-branes which wrap $\IP^1$s correspond to K\"ahler
blowups in four dimensions via the Penrose transform \ct{twistor-string}.
This can be interpreted as the back-reaction of the D1-branes in the
twistor space, which is precisely our picture if the D1-branes are
identified with NC $U(1)$ instantons, which are in turn
gravitational instantons.

The above construction is also very similar to topological D-branes on NC
manifolds in \ct{kapustin} which can be understood
in terms of generalized complex geometry \ct{generalized-geometry}.
Especially, see section 6 of the first paper in \ct{kapustin} where Eq.(6)
corresponds to our \eq{sw-instanton} or \eq{cma-pleb}.
This coincidence might be expected at the outset since the
generalized complex geometry incorporates
symplectic structures as well as usual complex structures
(see the footnote \ref{general-geometry})
and the emergent gravity is essentially based on a NC deformation of
symplectic structures. We will more exploit this relation in \ct{future}.

\section{NC Self-duality and Twistor Space}

In section 4, we ignored derivative corrections whose explicit forms are given
in section 2 and 3. Furthermore we used the usual NC description with $\Phi =
0$, which is not background independent, i.e., $\theta$-dependent
\ct{seiberg}. As a result, we separately considered two kinds of NC
instantons; Nekrasov-Schwarz instantons \ct{nek-sch} and
localized instantons \ct{minwalla}. In particular, the SW map of
localized instantons generated by shift operators was shown to be trivial,
i.e., $F^{\pm}_{\mu\nu} =0$ \ct{hashimoto-ooguri}, not to probe the geometry
by localized instantons. (This may be an artifact of the semi-classical
approximation.) Therefore the background independent formulation of
NC gauge theory \ct{sw,seiberg} might be more effective to have
a unified description for all possible backgrounds
and to implement a possible effect of derivative corrections.

In this section, we will generalize the equivalence in section 4
using the background independent formulation of NC gauge theories
and show that self-dual electromagnetism in NC spacetime is equivalent to
self-dual Einstein gravity, uncovering many details in \ct{mine}.
In particular, we will discuss in detail the twistor space structure
inherent in the self-dual NC electromagnetism. As a great bonus, the
background independent formulation clearly reveals a picture that
the NC gauge theory and gravity correspondence may be understood as
a large $N$ duality.

To see this picture, consider the SW map \eq{sw-equiv} at $\CO(\kappa^2)$
for the background independent case with $\Phi = - B$ where $B_{\mu\nu} =
(1/\theta)_{\mu\nu}$:
\begin{equation}
\label{sw-gen}
\frac{1}{4 G_s} \int d^4 y (\widehat{F} - B)^2
= \frac{1}{4 G_s} \int d^4 x  \sqrt{\det{\mathrm{g}}} \;
{\mathrm{g}}^{\mu\lambda} {\mathrm{g}}^{\sigma\nu}
B_{\mu\nu}B_{\lambda\sigma}.
\end{equation}
Although the right hand side is neglecting derivative corrections,
we will now use the full NC field strength \eq{nc-f} to examine
the derivative corrections. Later on, we will use only the left hand
side which can be rewritten in terms of closed string variables only
as follows
\begin{eqnarray} \label{matrix-sw}
&& \frac{1}{4 G_s}\int d^4 y (\widehat{F}-B)_{\mu\nu}
\star   (\widehat{F}-B)^{\mu\nu} \xx
&=& -\frac{\pi^2}{g_s \kappa^2} g_{\mu\lambda} g_{\nu\sigma}
{\mathbf{Tr}}_{\mathcal{H}} [x^\mu,x^\nu][x^\lambda, x^\sigma]
\end{eqnarray}
where we made a replacement $\frac{1}{(2\pi)^2} \int \frac{d^4y}{\mathrm{Pf}\theta}
\leftrightarrow  \mathbf{Tr}_{\mathcal{H}}$ using the Weyl-Moyal map \ct{nc-review}.
The covariant, background-independent coordinates $x^\mu$ are
defined by \eq{cov-coord} and they are now operators on an
infinite-dimensional, separable Hilbert space $\mathcal{H}$,
which is the representation space of the Heisenberg
algebra \eq{nc-spacetime}. The NC gauge symmetry in Eq.(\ref{matrix-sw}) then acts
as unitary transformations on $\mathcal{H}$, i.e.,
\be \la{nc-symmetry}
x^\mu \rightarrow {x^{\prime}}^\mu = U x^\mu U^\dagger.
\ee
This NC gauge symmetry $U_{\rm{cpt}}(\mathcal{H})$ is so large that
$U_{\rm{cpt}}({\mathcal{H}}) \supset U(N) \;(N \rightarrow \infty)$ \cite{harvey}.
In this sense the NC gauge theory in Eq.(\ref{matrix-sw})
is essentially a large $N$ gauge theory.
Note that the second expression in Eq.\eq{matrix-sw} is a large $N$ version of
the IKKT matrix model which describes the nonperturbative dynamics of type IIB
string theory \ct{ikkt}.

Now let us apply the gauge equivalence in Eq.\eq{star-equiv}
or Eq.\eq{zotov-star} for the adjoint action of $x^\mu$ with respect to
star product:
\bea \la{adjoint-star}
[x^\mu, \what{f}]_\star &=& 2 \hbar D(\hbar)^{-1} \Bigl(
\a^{\mu\nu}(x) \frac{\p f}{\p x^\nu} \Bigr) + \CO(\hbar^3) \xx
&\approx& 2\hbar \a^{\mu\nu}(x) \frac{\p f}{\p x^\nu} +  \CO(\hbar^3)
\eea
where $\what{f} \equiv D(\hbar)^{-1}(f)$. If $\what{f}=x^\nu$, we recover
Eq.\eq{esw-deformation}.

Beyond the semi-classical approximation, Eq.\eq{adjoint-star} is not reduced
to usual vector fields since there are infinitely many derivatives as shown in
Eqs.\eq{zotov-star} and \eq{diff-d}. But it is important to notice
the following properties (see the footnote \ref{poisson}), where we use the
operator notation using
the Weyl-Moyal map \eq{star-product} for definiteness
\bea \la{derivation1}
&& [\what{x}^\mu, \what{f}\,\what{g}] = [\what{x}^\mu, \what{f}]\what{g}
+ \what{f} [\what{x}^\mu, \what{g}], \\
\la{derivation2}
&& [\what{x}^\mu, [\what{x}^\nu, \what{f}]] -
[\what{x}^\nu, [\what{x}^\mu, \what{f}]] = [[\what{x}^\mu, \what{x}^\nu],
\what{f}]].
\eea
These properties show that ${\rm ad}_{x^\mu} \equiv [x^\mu, \cdot \,]_\star$
generally satisfy the property of vector fields or Lie
derivatives even after quantum deformation.
(Also note that $D_\mu \equiv -i B_{\mu\nu} x^\nu$ is a covariant derivative
in NC gauge theory.)
Indeed this kind of vector fields was already defined in terms of twisted
diffeomorphisms \ct{nc-gravity,nc-gravity2,szabo},
where a vector field on $\IR^d$ becomes a higher-order
differential operator acting on fields in $\CA$.
We thus see that Eq.\eq{adjoint-star} defines {\it generalized}
vector fields according to Eqs.\eq{derivation1} and \eq{derivation2}.

The appearance of NC gravity framework in our context might be
anticipated. We observed in section 3 that the emergent gravity is
related to a NC deformation of diffeomorphism symmetry \eq{darboux}.
In other words, the Darboux theorem in symplectic geometry can be regarded as
the equivalence principle in general relativity, but in general we need a
NC version of the equivalence principle since we now live in
the NC phase space \eq{nc-spacetime}. Actually, this is an underlying
principle of NC gravity \ct{nc-gravity}. We will more clarify in \ct{future}
the emergent gravity from the viewpoint of NC gravity.

Let us now return to the semi-classical limit $\CO(\hbar)$.
In this limit,
\be \la{vector-classical}
[x^\mu, f]_\star \approx i \hbar \theta^{\a\b}
\frac{\p x^\mu}{\p y^\a} \frac{\p f}{\p y^\b} = i\hbar \{ x^\mu ,f \}
\equiv V^\mu [f].
\ee
As expected, the adjoint action of $x^\mu$ with respect to
star product reduces to a vector field $V_\mu \in T{\cal M}$ on some emergent
four manifold ${\cal M}$.
This is precisely the limit in section 4 that NC electromagnetism
reduces to Einstein gravity for the (A)SD sectors.
Note that NC gauge fields $\widehat{A}_\mu(y)$ are in general arbitrary,
so they generate arbitrary vector fields $V^\mu \in T{\cal M}$
according to the map \eq{vector-classical} and $[x^\mu, f]_\star = i \theta^{\mu\nu}
\partial_\nu f$ when $\widehat{A}_\mu = 0$.
One can easily check that
\be \la{lie-bracket}
({\rm ad}_{x^\mu}\, {\rm ad}_{x^\nu}-{\rm ad}_{x^\nu}\,{\rm ad}_{x^\mu})[f]
={\rm ad}_{[x^\mu,x^\nu]_\star}[f] = [V^\mu,V^\nu][f]
\ee
where the right-hand side is defined by the Lie bracket between vector
fields in $T{\cal M}$. Note that the gauge transformation \eq{nc-symmetry}
naturally induces coordinate transformations of frame fields
\be \la{vector-tr}
V^{\mu \alpha} \rightarrow {V^\prime}^{\mu \alpha}
= \frac{\partial {y^\prime}^{\alpha}}{\partial y^\beta} V^{\mu \beta}.
\ee
This leads to a consistent result \ct{mine} that the gauge equivalence due to
\eq{nc-symmetry} corresponds to the diffeomorphic equivalence between the
frame fields $V^\mu$.

Let us look for an instanton solution of Eq.(\ref{matrix-sw}).
Since the instanton is a Euclidean solution with a finite action,
the instanton configuration should approach a pure gauge at
infinity. Our boundary condition is ${\widehat F}_{\mu\nu}
\to 0$ at $|y| \to \infty$ for the instanton configuration.
Thus one has to remove parts due to backgrounds
from the action in Eq.(\ref{matrix-sw}).
One can easily achieve this by defining the self-duality equation as
follows \cite{mine}
\bea \la{sde-matrix}
&& {\rm ad}_{[x^\mu,x^\nu]_\star} =
\pm \half \varepsilon_{\mu\nu\lambda\sigma} \;
{\rm ad}_{[x^\lambda, x^\sigma]_\star} \xx
\Leftrightarrow &&
[V_\mu,V_\nu] =
\pm \half \varepsilon_{\mu\nu\lambda\sigma} [V_\lambda, V_\sigma],
\eea
where we used Eq.\eq{lie-bracket}.
From the above definition, it is obvious that the constant part in
$[x^\mu,x^\nu]_\star = -i (\theta({\widehat F}-B)
\theta)^{\mu\nu}$, i.e. $i \theta^{\mu\nu}$, can be dropped.
Note that, for nondegenerate $\theta^{\mu\nu}$'s, the first self-duality
equation in Eq.\eq{sde-matrix} reduces to Eq.\eq{nc-self-dual}.
An advantage of background independent formulation is that
Eq.\eq{sde-matrix} holds for an arbitrary non-degenerate $\theta^{\mu\nu}$
and there is no need to specify a background.

It is obvious that the vector fields preserve the volume
form $\vare_4$, i.e., ${\cal L}_{V_\mu} \vare_4 = 0$,
where ${\cal L}_{V_\mu}$ is the Lie derivative along $V_\mu$
since all the vector fields $V_\mu$ are divergence free,
i.e. $\p_\a V_{\mu}^\a=0$.
Incidentally, this is simply the Liouville theorem in
symplectic geometry \ct{books}.
In consequence, instanton configurations are mapped to
the volume preserving diffeomorphism, $SDiff({\cal M})$, satisfying
Eq.\eq{sde-matrix}.

So we arrive at the result of Ashtekar {\it et al.} \cite{ashtekar}.
Their result is summarized as follows \ct{joyce}.
Let ${\cal M}$ be an oriented 4-manifold and let $V_\mu$ be
vector fields on ${\cal M}$ forming an oriented basis for $T{\cal M}$.
Then $V_\mu$ define a conformal structure $[G]$ on ${\cal M}$.
Suppose that $V_\mu$ preserve a volume form on ${\cal M}$
and satisfy the self-duality equation
\begin{equation} \label{ashtekar}
[ V_\mu, V_\nu ] = \pm \frac{1}{2} {\varepsilon}_{\mu\nu\lambda\sigma}
[ V_\lambda, V_\sigma ].
\end{equation}
Then $[G]$ defines an (anti-)self-dual and Ricci-flat metric.

The (inverse) metric determined by the vector fields in
Eq.\eq{vector-classical} is then given by \ct{ashtekar}
\be \label{sd-metric}
G^{\a \b} = {\det V}^{-1}V_\mu^\alpha V_\nu^\beta \delta^{\mu\nu},
\ee
where the background spacetime metric was taken as
$g_{\mu\nu}= \delta_{\mu\nu}$ for simplicity.

Motivated by the similarity of Eq.\eq{ashtekar} to the self-duality equation
of Yang-Mills theory, Mason and Newman showed \cite{mason-newman}
that, if we have a reduced Yang-Mills theory
where the gauge fields take values in the Lie algebra of $SDiff({\cal M})$,
which is exactly the case for the action \eq{matrix-sw} through
the map \eq{vector-classical}, Yang-Mills instantons are actually equivalent
to gravitational instantons.\footnote{In their approach, there exists a
tetrad freedom which was referred to as a (metric-preserving) gauge transformation.
We also showed that the NC $U(1)$ gauge symmetry \eq{nc-symmetry} appears
as the diffeomorphic equivalence \eq{vector-tr} between
the metrics on $\CM$.}
We showed that this is the case for NC electromagnetism.
See also related works \ct{qhpark,ward}.

Since the second expression in Eq.\eq{matrix-sw} is the bosonic part of
the IKKT matrix model \ct{ikkt}, our current result is consistent with the claim
that the IKKT matrix model is a theory of gravity (or type IIB string
theory). See also recent works \ct{ikkt-gravity} addressing this issue
directly from IKKT matrix model.
In addition the result in \ct{kita-naga} obviously indicates the existence
of 4-dimensional massless gravitons in NC gauge theory, which supports our
claim about the emergent gravity from NC electromagnetism.\footnote{We are grateful to
S. Nagaoka for drawing our attention to their paper.}

It is {\it a priori} not obvious that the self-dual electromagnetism in NC
spacetime is equivalent to the self-dual Einstein gravity.
Therefore it should be helpful to have explicit nontrivial examples
to appreciate how it works. It is not difficult to find them from
Eq.\eq{ashtekar}, which was already done for the Gibbons-Hawking metric
\ct{gibb-hawk} in \ct{joyce} and for the real heaven solution \ct{real-heaven}
in \ct{real-heaven2}.

The Gibbons-Hawking metric \ct{gibb-hawk} is a general class of self-dual,
Ricci-flat metrics with the triholomorphic $U(1)$ symmetry
which describes a particular class of ALE and asymptotically
locally flat (ALF) instantons.
Let $(a_i, U),\; i=1,2,3,$ are smooth real functions on $\IR^3$
and define $V_i = - a_i \frac{\partial}{\partial \tau}
+ \frac{\partial}{\partial x^i}$ and $V_4 = U \frac{\partial}{\partial \tau}$,
where $\tau $ parameterizes circles and the Killing vector
$\p / \p \tau$ generates the triholomorphic $U(1)$ symmetry.
Eq.(\ref{ashtekar}) then becomes the equation $ \nabla U +
\nabla \times \vec{a} = 0$ and the metric whose inverse
is \eq{sd-metric} is given by
\be \label{gh-metric}
ds^2 = U^{-1}(d\tau + \vec{a} \cdot d\vec{x})^2 + U d\vec{x} \cdot d\vec{x},
\ee
where $\vec{x} \in \IR^3$.

The real heaven metric \ct{real-heaven} describes four dimensional
hyper-K\"ahler manifolds with a rotational Killing symmetry which is
also completely determined by one real scalar field \ct{bakas}.
The vector fields $V_\mu$ in this case are given by \ct{real-heaven2}
\bea \la{heaven-vector}
&& V_1 = \frac{\p}{\p x^1} - \p_2 \psi \frac{\p}{\p \tau} \xx
&& V_2 = \frac{\p}{\p x^2} + \p_1 \psi \frac{\p}{\p \tau} \\
&& V_3 = e^{\psi/2} \left ( \sin \Bigl(\frac{\tau}{2} \Bigr) \frac{\p}{\p x^3}
+ \p_3 \psi \cos \Bigl(\frac{\tau}{2} \Bigr) \frac{\p}{\p \tau}  \right) \xx
&& V_4 = e^{\psi/2} \left( \cos \Bigl(\frac{\tau}{2}\Bigr) \frac{\p}{\p x^3}
- \p_3 \psi \sin \Bigl(\frac{\tau}{2} \Bigr) \frac{\p}{\p \tau} \right)
\nonumber
\eea
where the rotational Killing vector is given by $c_i \p_i \psi \p/\p \tau$
with constants $c_i \; (i=1,2)$ and the function $\psi$
is independent of $\tau$. Eq.(\ref{ashtekar}) is then equivalent to
the three-dimensional continual Toda equation $(\p_1^2 + \p_2^2) \psi
+ \p_3^2 e^\psi = 0$ and the metric is determined by Eq.\eq{sd-metric} as
\be \label{heaven-metric}
ds^2 = (\p_3 \psi)^{-1}(d\tau +  a^i d x^i)^2 + (\p_3 \psi)
( e^\psi dx^i dx^i + dx^3 dx^3)
\ee
where $a^i = \varepsilon^{ij} \p_j \psi$.

The canonical structures, in particular, complex and K\"ahler structures,
of the self-dual system \eq{ashtekar}, have been fully studied in a
beautiful paper \ct{beautiful}. The arguments in \ct{beautiful} are
essentially the same as ours leading to Eq.\eq{cma-pleb}. It was also shown
there how the Pleba\~nski's heavenly equations \ct{plebanski} can be derived
from Eq.\eq{ashtekar}. It should be interesting to recall
\ct{husain} that Eq.\eq{ashtekar} can be reduced to the $sdiff(\Sigma_g)$
chiral field equations in two dimensions, where $sdiff(\Sigma_g)$ is the area
preserving diffeomorphisms of a Riemann surface of genus $g$.

Now we will study in detail the structure of twistor space inherent
in Eq.\eq{ashtekar}. Define holomorphic vector fields $\CV$ and $\CW$ locally by
\be \la{hol-vector}
\CV = V_2 + iV_1 = f_i \frac{\p}{\p z_i}, \quad
\CW = V_4 + iV_3 = g_i \frac{\p}{\p z_i},
\ee
where $f_i, g_i \;(i=1,2)$ are complex functions on $\CM$.
In terms of these vector fields, Eq.\eq{ashtekar} reduces to the following
triple
\be \la{triple}
[\CV,\bar{\CW}] = 0, \qquad [\CV,\bar{\CV}] = [\CW,\bar{\CW}].
\ee
Substituting \eq{hol-vector} into \eq{triple}, we find that the equations
are satisfied identically if
\be \la{hol-function}
\frac{\p f_i}{\p \bar{z}_j} = \frac{\p g_i}{\p \bar{z}_j} = 0.
\ee
So we can construct a hypercomplex (or hyper-K\"ahler) structure on $\CM$
locally out of four holomorphic functions $f_i$ and $g_i$,
or globally out of two holomorphic vector fields \ct{joyce}.

If we introduce the following $\IP^1$-family of vector fields parameterized by
$t \in \IP^1$,
\be \la{family-vector}
\CL = \CV + t \CW, \qquad \CN = \bar{\CW} - t \bar{\CV},
\ee
the self-dual Einstein equations \eq{ashtekar} are more compactly written as
\be \la{lax}
[\CL,\CN] = 0
\ee
with the volume preserving constraint
\be \la{sdiff}
\CL_{\CL} \vare_4 = 0 = \CL_{\CN} \vare_4.
\ee
Eq.\eq{lax} can be interpreted as a Lax pair form of curved self-dual spacetime.

It is easy to see that Eq.\eq{triple} defines a hyper-K\"ahler structure on
$\CM$. We construct a $t$-dependent two-form $\Psi(t)$ on $\CM$ by contracting
the volume form $\vare_4$ by $\CL$ and $\CN$
\be \la{hol-twistor}
\Psi(t) = \vare_4 (\, \cdot \,, \, \cdot \,, \CL, \CN) = \w + it \Omega + t^2
\bar{\w}
\ee
where
\bea \la{hyper-3}
&& \w = \vare_4 (\, \cdot \,, \, \cdot \,, \CV, \bar{\CW}), \xx
&& i \Omega  = \vare_4 (\, \cdot \,, \, \cdot \,, \CW, \bar{\CW})
- \vare_4 (\, \cdot \,, \, \cdot \,, \CV, \bar{\CV}) \\
&& \bar{\w} = \vare_4 (\, \cdot \,, \, \cdot \,, \bar{\CV}, \CW). \nonumber
\eea
Note that we need the property \eq{lax} to make sense of Eq.\eq{hol-twistor}.
The two-form $\Psi(t)$ in \eq{hol-twistor} is an exact analogue of
Eq.\eq{deformation} and the resulting consequences are exactly parallel to
section 4. Nevertheless it will be useful to understand parallel arguments with
section 4 for this purely geometrical setting since it seems to be very
powerful for later applications.

It is straightforward to prove (see Eq.(8) in \ct{beautiful})
using the Cartan's homotopy formula $\CL_X = \iota_X d + d
\iota_X$ that $\Psi(t)$ is closed, i.e., $d \Psi(t) = 0$. We see from the
proof that Eq.\eq{sdiff} is analogous to the Bianchi identity.
We can thus define on $\CM$ the three non-degenerate symplectic forms
\be \la{3-kahler}
\w^a = \Bigl( \w^1= -\frac{i}{2} (\w - \bar{\w}), \w^2 = \half (\w +
\bar{\w}), \w^3 = - \Omega \Bigr).
\ee
$\w^a$ are also three K\"ahler forms compatible with three complex structures
on $\CM$ (see section IV. A and B in \ct{beautiful}) and thus define the
hyper-K\"ahler structure on $\CM$. Therefore any metric defined by
Eq.\eq{ashtekar} with the constraint \eq{sdiff} should be hyper-K\"ahler,
as also shown in section 4.

Since the two-form $\Psi(t)$ is closed and degenerate for any $t \in \IP^1$,
one can introduce holomorphic coordinates in a natural fashion via the Darboux
theorem on each coordinate chart on $\IP^1$ such that
$\Psi(t)$ is a holomorphic (2,0)-form on the local chart.
For example, Eq.\eq{symplectic-form} in a neighborhood of $t=0$
(the south pole of $\IP^1$) and Eq.\eq{symplectic-form2} in a neighborhood of
$t=\infty$ (the north pole of $\IP^1$).
But they can be consistently patched
along the equator of $\IP^1$ in such a way that the total space $\CZ$, the
twistor space, including $\IP^1$ becomes a three-dimensional complex manifold
as we explained in section 4.

We know from Eq.\eq{hyper-3} that
$\Omega$ is rank 4 while $\w$ and $\bar{\w}$ are both rank 2.
As a direct consequence, we immediately get Eq.\eq{cma-pleb} \ct{beautiful}.
In terms of local coordinates, it reduces to the complex Monge-Amp\`ere
equation \ct{yau} or the Pleba\~nski equation \ct{plebanski}.
Since $\Omega = dx^1 \wedge dx^2 + dx^3 \wedge dx^4 + \zeta(B+F)$
in Eq.\eq{hyper-3} can always serve as a symplectic form on
both coordinate charts (note that it is rank 4), two sets of coordinates
at $t=0$ and $t=\infty$ should be related to each other
by a canonical transformation (where we refer the canonical
transformation in a more general sense).
A beautiful fact was shown in \ct{beautiful} that the canonical transformation
between them is generated by the K\"ahler potential appearing in the complex
Monge-Amp\`ere equation or the Pleba\~nski equation. In other words, the
K\"ahler potential is a generating function of canonical transformations or
a transition function of three-dimensional complex manifold $\CZ$ as a
holomorphic vector bundle \ct{atiyah}.

Finally we would like to discuss an interesting fact that
Eq.\eq{ashtekar} can be reduced to $sdiff(\Sigma_g)$ chiral field
equations in two dimensions \ct{husain}.
We will directly show using canonical transformations that
the Husain's equation \ct{husain} (where we denote $\Lambda_x =
\p_x \Lambda, \; \Lambda_{xq} = \p_x \p_q \Lambda $, etc.)
\be \la{husain}
\Lambda_{xx} + \Lambda_{yy} + \Lambda_{xq}\Lambda_{yp}-
\Lambda_{xp}\Lambda_{yq} = 0
\ee
is equivalent to the first heavenly equation \ct{plebanski},
which is a governing equation of self-dual Einstein gravity.
This implies that the self-dual system \eq{ashtekar} is deeply related
to two dimensional $SU(\infty)$ chiral models \ct{qhpark,ward}.
An interesting implication of this connection
will be briefly discussed in next section.

Although the first heavenly equation was also obtained in \ct{husain}
by a different reduction of Eq.\eq{ashtekar}, an explicit canonical
transformation between them was not available there. In the course of
derivation, we will find an interesting symplectic structure of
Eq.\eq{ashtekar} which was also noticed in \ct{tafel}
from a different approach. Ours is more straightforward.

By complex coordinates, $u= x+iy, \; v = q+ ip$, Eq.\eq {husain} reads as
\be \la{husain-complex}
\Lambda_{u\bar{u}}-(\Lambda_{uv} \Lambda_{\bar{u}\bar{v}}
- \Lambda_{u\bar{v}} \Lambda_{\bar{u}v}) = 0.
\ee
We will now apply a similar strategy as the Appendix in \ct{grant}.
Define two functions $A = \Lambda_u$ and $B = \Lambda_{\bar{u}}$.
Eq.\eq{husain-complex} is then equivalent to
\bea \label{husain1}
&& A_{\bar{u}} - (A_{v} B_{\bar{v}}- A_{\bar{v}} B_{v}) = 0 \\
\label{husain2}
&& A_{\bar{u}} = B_{u}.
\eea
Instead of looking on $A$ as a function of $(u,v,\bar{u},\bar{v})$,
we take $A$ as a coordinate and look on $f \equiv \bar{u}$ and $g \equiv B$
as functions of $(\xi^1 \equiv A, \xi^2 \equiv u, {\tilde \xi}^1 \equiv v,
{\tilde \xi}^2 \equiv \bar{v})$.
This is a canonical transformation which is well-defined
as long as $A_{\bar{u}} \neq 0$.

It is convenient to denote coordinates by $\xi^A, {\tilde \xi}^A, \; A=1,2$
for compact notation and to use the antisymmetric tensors $\epsilon^{AB}$ and
$\epsilon^{\tilde{A}\tilde{B}}$ to raise indices in a standard way,
e.g. $\xi^A = \epsilon^{AB} \xi_B,  {\tilde \xi}^A =
\epsilon^{\tilde{A}\tilde{B}}{\tilde \xi}_B $.
It is easy to show that, after the above coordinate transformation,
Eq.\eq{husain1} and Eq.\eq{husain2} are transformed to (after applying a
series of chain rules)
\be \la{canon1}
\epsilon^{\tilde{A}\tilde{B}} \partial_{\tilde{A}} f \partial_{\tilde{B}} g = 1
\ee
and
\be \la{canon2}
\epsilon^{AB} \partial_A f \partial_B g = 1,
\ee
respectively. Thus the Husain's equation is reduced to the two Poisson bracket relations
which relate $f$ and $g$ to $\xi^A$ and ${\tilde \xi}^A$ by canonical
transformations.

One can show that Eqs.\eq{canon1} and \eq{canon2} lead to the result
\ct{tafel} that there is a function $K$ such that
\be \la{tafel-omega}
\partial_{A} f \partial_{\tilde{B}} g -  \partial_{\tilde{B}} f \partial_{A} g
= \partial_{A} \partial_{\tilde{B}} K.
\ee
Using some relation between Poisson brackets (Eq.(11) in \ct{tafel}),
we arrive at the result that $K$ satisfies the Pleba\~nski equation
\be \la{1st-heaven}
\epsilon^{\tilde{A}\tilde{B}}
\partial_{A} \partial_{\tilde{A}} K \; \partial_{B} \partial_{\tilde{B}} K
= \epsilon_{AB}.
\ee
This completes the proof that Eq.\eq{husain} is equivalent to the first
heavenly equation \eq{1st-heaven}.

\section{Discussion}

Let us briefly recapitulate our main results.
A basic reason for the emergent gravity from NC spacetime is that
the $\Lambda$-symmetry \eq{lambda-symmetry} can be regarded as a par with
$Diff(M)$, which results from the Darboux theorem in
symplectic geometry. The spontaneous symmetry breaking \eq{spacetime-vacuum}
also comes into play for the emergent gravity.
In general the emergent gravity needs to incorporate NC deformations of
the diffeomorphism symmetry since it should be defined
on NC spacetime \eq{nc-spacetime}. In this context,
the gauge equivalence \eq{star-equiv} in deformation
quantization might be interpreted as a {\it quantum equivalence principle}.

We have derived the exact SW maps with derivative corrections in two ways;
from the SW equivalence \eq{sw-sw-equivalence} and
from the deformation quantization \eq{star-equiv}.
But they should be the same since the SW equivalence
\eq{general-sw-equivalence} is the equivalent statement as
the gauge equivalence \eq{star-equiv} as we showed in the
semi-classical limit. It should be interesting, in its own right,
to explicitly check the consistency between two different approaches
for the derivative corrections.

We showed in section 4 and 5 that the self-dual Einstein gravity is
emerging from self-dual NC electromagnetism neglecting derivative
corrections, i.e., defined with the Poisson bracket \eq{poisson-bracket}.
We thus expect that the derivative corrections give rise to higher
order gravity, e.g., $R^2$ gravity.
It should be important to precisely determine the form of the higher order gravity.
Since the emergent gravity is in general a full quantum deformation of $Diff(M)$,
it might modestly be identified with a NC gravity \ct{nc-gravity}, as we
argued in section 5. If this is the case, the SW maps in section 2
and 3 including derivative corrections may be related to those of NC
gravity. Our construction in section 4 and 5 also implies that we
need a NC
deformation of twistor space \ct{nc-twistor} to describe general
nonlinear gravitons in NC gravity.

Recently, it was found \ct{twist-symm} that NC field theory is invariant
under the twisted Poincar\'e symmetry where the action of generators is now
defined by the twisted coproduct in the deformed Hopf algebras.
We think that the twisted Poincar\'e symmetry, especially the deformed Hopf
algebra and quantum group structures, will be important to
understand the NC field theory and gravity correspondence since
underlying symmetries are always an essential guide for physics.
Actually this symmetry plays a prominent role to construct
NC gravity \ct{nc-gravity,szabo}

Unlike the homogeneous background \eq{spacetime-vacuum},
there could be an inhomogeneous condensation of $B$-fields in a
vacuum. In this case, we expect a nontrivial curved spacetime
background, e.g., a Ricci-flat Einstein manifold instead of flat $\IR^4$
and we need a quantization on general symplectic (or Poisson)
manifolds \ct{szabo}. Our approach suggests an intriguing picture for an
inhomogeneous background, for example, specified by
\be \la{curved-spacetime}
\langle B^\prime_{\mu\nu}(x) \rangle_{\rm{vac}}
= ({\theta^\prime}^{-1})_{\mu\nu}(x).
\ee
One may regard $B^\prime_{\mu\nu}(x)$ as coming from an inhomogeneous
gauge field condensation on a constant $B_{\mu\nu}$ background, say,
$B^\prime_{\mu\nu}(x) = (B+ F_{\rm{back}}(x))_{\mu\nu}$.
For instance, if $F_{\rm{back}}(x)$ is an instanton, our result implies that
the vacuum manifold \eq{curved-spacetime} is a Ricci-flat K\"ahler manifold.
From NC gauge theory point of view, this corresponds to the description
of NC gauge theory in instanton backgrounds \ct{lee-yang}.
Therefore the NC gauge theory with nonconstant NC parameters
${\theta^\prime}^{\mu\nu}(x)$ may be interpreted as that defined
by the usual Moyal star product \eq{star-product}
but around a nonperturbative solution described by $F_{\rm{back}}(x)$.
The gravity picture in this case corresponds to a (perturbative)
NC gravity on a curved manifold. It will be interesting to see
whether this reasoning can shed some light on NC gravity.

Recently we suggested in \ct{ahrenshoop} a very simple toy model
for emergent gravity. We claimed that (2+1)-dimensional NC field theory
for a real scalar field in large NC limit $\theta \to \infty$
is equivalent to two dimensional string theory via $c=1$ Matrix
model. See \ct{sjrey} for a field theory discussion from this aspect.
This claim is based on the well-known relation \ct{sun-sdiff}
\be \la{sun-sdiff}
\mathrm{Real \; field \; on \; NC} \; \IR^2 \;
(\mathrm{or} \; \Sigma_g) \Longleftrightarrow  N \times N \; \mathrm{Hermitian \;
  matrix} \; \mathrm{at} \; N \to \infty,
\ee
where $\Sigma_g$ is a Riemann surface of genus $g$ which can be quantized via
deformation quantization. In two dimensions, a symplectic 2-form $\w$
is a volume form and Hodge-dual to a real function. So symplectomorphism is
equal to area-preserving diffeomorphism (APD).
(In higher dimensions, symplectomorphism is much smaller
than volume-preserving diffeomorphism.)
We observed in Eq.\eq{nc-gauge-tr} that symplectomorphism can be
identified with NC gauge symmetry.
In two dimensions, we thus have the relation: Symplectomorphism = APD = NC
gauge symmetry. So, if there is a NC field theory which is gauge invariant,
the NC field theory is then APD invariant and thus
we expect an emergent gravity in two dimensions from this NC field theory.

We can infer the nature of two-dimensional emergent gravity from
four-dimensional case.
Noting that the electromagnetic 2-form $F$ acts as a deformation of the
symplectic (or K\"ahler) structure, it is natural
to guess that a real scalar field plays the same role in two
dimensions. Since the K\"ahler potential behaves as a
generating function of canonical transformations as we observed in
section 5, it is also plausible that the real scalar field is a
generating function of APDs and acts as a K\"ahler potential. We hope to discuss this
interesting correspondence in a separate publication.

In section 5, we showed that the Husain's equation \eq{husain} is equivalent
to the first heavenly equation \eq{1st-heaven}. Here we note that Eq.\eq{husain}
is the $SU(N)$ self-dual Yang-Mills equation in the limit $N \to \infty$
\ct{ym-instanton}, which implies that $SU(N \to \infty)$ Yang-Mills instantons
are gravitational instantons too.
This interesting fact is also coming from the relation $(\Longleftarrow)$
in \eq{sun-sdiff} since the gauge fields in $SU(N)$ Yang-Mills theory
on $\IR^4$ are all $N \times N$ Hermitian matrices
and thus they can be mapped to real scalar fields on a
six-dimensional space $\IR^4 \times \Sigma_g$. This seems to imply that
the AdS/CFT duality \ct{ads-cft} might be deeply related to the NC field
theory and gravity correspondence.

In order to more extensively understand the nature of emergent
gravity, it is useful to consider couplings with matter fields. To
do this, we need to know the SW maps for currents and energy-mometum
tensors for matter fields. These were obtained in \ct{bly} at leading order.
It turned out \ct{rivelles,ban-yang} that the gravitational coupling with
matter fields is not universal unlike as general relativity. It deserves to
ask more study, especially for experimental verifications.

The emergent gravity from NC gauge theory discussed in this paper
may have interesting implications to string theory and black-hole physics.
We briefly discuss possible implications citing relevant literatures.

It was argued \ct{panic} that tachyon condensation at the fixed points of noncompact
nonsupersymmetric orbifolds, e.g. $\IC^2/\Gamma$, drives these orbifolds
to flat space or supersymmetric ALE spaces. But ALE spaces are $U(1)$
instantons in flat NC $\IR^4$ \ct{sty,ys}. Does it imply
that the closed string tachyon condensation can be understood as an
open string tachyon condensation ?
The picture in \ct{koji-seiji} may be useful for this problem.

Microscopic black hole entropy in string theory \ct{black-hole} was derived
by counting the degeneracy of BPS soliton bound states,
mostly involved with instanton moduli space. If we simply assume that the
instanton moduli space is coming from NC $U(1)$ instantons, then the counting
of the degeneracy is just the counting of all possible hyper-K\"ahler geometries
inside the black hole horizon, according to our picture.
This is very reminiscent of the Mathur's program
for black hole entropy \ct{mathur}.

We showed that the equivalence between NC $U(1)$ instantons and
gravitational instantons could be beautifully understood in terms of
the twistor space. We think that the equivalence and its twistor space
structure should have far-reaching applications to Nekrasov's instanton
counting \ct{instanton-counting}
and topological strings for crystal melting \ct{top-string}.

\hspace{1cm}

{\bf Acknowledgments} We would like to thank Ilka Agricola, Harald Dorn,
J\"urg K\"appeli, and Fabian Spill for helpful discussions.
We are also grateful to Kyung-Seok Cha, Bum-Hoon Lee, and Jongwon
Lee for collaboration about derivative corrections at the very early
stage. Some of the results in this paper were reported at the
Ahrenshoop workshop \ct{ahrenshoop}. We would like to thank the organizers
for their kind invitation and a very enjoyable meeting. This work
was supported by the Alexander von Humboldt Foundation.

\newpage


\nc{\npb}[3]{Nucl. Phys. {\bf B#1} (#2) #3}

\nc{\plb}[3]{Phys. Lett. {\bf B#1} (#2) #3}

\nc{\prl}[3]{Phys. Rev. Lett. {\bf #1} (#2) #3}

\nc{\prd}[3]{Phys. Rev. {\bf D#1} (#2) #3}

\nc{\ap}[3]{Ann. Phys. {\bf #1} (#2) #3}

\nc{\prep}[3]{Phys. Rep. {\bf #1} (#2) #3}

\nc{\epj}[3]{Eur. Phys. J. {\bf #1} (#2) #3}

\nc{\ptp}[3]{Prog. Theor. Phys. {\bf #1} (#2) #3}

\nc{\rmp}[3]{Rev. Mod. Phys. {\bf #1} (#2) #3}

\nc{\cmp}[3]{Commun. Math. Phys. {\bf #1} (#2) #3}

\nc{\mpl}[3]{Mod. Phys. Lett. {\bf #1} (#2) #3}

\nc{\cqg}[3]{Class. Quant. Grav. {\bf #1} (#2) #3}

\nc{\jhep}[3]{J. High Energy Phys. {\bf #1} (#2) #3}

\nc{\atmp}[3]{Adv. Theor. Math. Phys. {\bf #1} (#2) #3}

\nc{\hepth}[1]{{\tt hep-th/{#1}}}


\end{document}